\documentclass[10pt,twocolumn,letterpaper]{article}

\usepackage[pagenumbers]{cvpr} 

\usepackage{graphicx}
\usepackage{amsmath}
\usepackage{amssymb}
\usepackage{booktabs}

\usepackage{lipsum}
\usepackage{blindtext}
\usepackage{multirow}
\usepackage{pifont}
\newcommand{\cmark}{\ding{51}}%
\newcommand{\xmark}{\ding{55}}%
\usepackage[table]{xcolor}
\definecolor{maroon}{cmyk}{0,0.87,0.68,0.32}
\definecolor{myyellow}{RGB}{218, 160, 109}
\definecolor{brickred}{rgb}{0.8, 0.25, 0.33}
\definecolor{brandeisblue}{rgb}{0.0, 0.44, 1.0}
\definecolor{applegreen}{rgb}{0.55, 0.71, 0.0}
\definecolor{aogreen}{rgb}{0.0, 0.5, 0.0}

\definecolor{b}{RGB}{47, 114, 173}  
\definecolor{r}{RGB}{199, 100,  38}
\definecolor{g}{RGB}{70, 155, 118}
\definecolor{m}{RGB}{193, 126, 165}
\definecolor{y}{RGB}{239, 227,  98}
\definecolor{c}{RGB}{110, 179, 228}
\definecolor{k}{RGB}{20, 20, 20}
\definecolor{turquoise}{cmyk}{0.65,0,0.1,0.3}
\definecolor{purple}{rgb}{0.65,0,0.65}
\definecolor{dark_green}{rgb}{0, 0.5, 0}
\definecolor{orange}{rgb}{0.8, 0.6, 0.2}
\definecolor{red}{rgb}{0.8, 0.2, 0.2}
\definecolor{darkred}{rgb}{0.6, 0.1, 0.05}
\definecolor{blueish}{rgb}{0.0, 0.3, .6}
\definecolor{light_gray}{rgb}{0.7, 0.7, .7}
\definecolor{pink}{rgb}{1, 0, 1}
\definecolor{greyblue}{rgb}{0.25, 0.25, 1}
\definecolor{orgred}{rgb}{1.0, 0, 0}

\usepackage{array}
\newcolumntype{C}[1]{>{\centering\let\newline\\\arraybackslash\hspace{0pt}}m{#1}}

\usepackage{algorithm}
\usepackage{algorithm}
\usepackage{algorithmic}
\usepackage{listings}
\usepackage{enumitem}

\usepackage[pagebackref=true,breaklinks=true,colorlinks,citecolor=blue,linkcolor=orgred,bookmarks=false]{hyperref}

\usepackage[capitalize]{cleveref}
\crefname{section}{Sec.}{Secs.}
\Crefname{section}{Section}{Sections}
\Crefname{table}{Table}{Tables}
\crefname{table}{Tab.}{Tabs.}

\def\cvprPaperID{2106} 
\def\confName{CVPR}
\def\confYear{2022}

\begin{document}

\title{MAXIM: Multi-Axis MLP for Image Processing}



\author{
Zhengzhong Tu$^{1,2}$\thanks{Work done during an internship at Google.} \qquad
Hossein Talebi$^{1}$ \qquad
Han Zhang$^{1}$ \qquad
Feng Yang$^{1}$ \\
Peyman Milanfar$^{1}$ \qquad
Alan Bovik$^{2}$ \qquad
Yinxiao Li$^{1}$ \\
$^1$ Google Research \qquad  $^2$ University of Texas at Austin\\
}

\maketitle

\begin{abstract}
Recent progress on Transformers and multi-layer perceptron (MLP) models provide new network architectural designs for computer vision tasks. Although these models proved to be effective in many vision tasks such as image recognition, there remain challenges in adapting them for low-level vision. The inflexibility to support high-resolution images and limitations of local attention are perhaps the main bottlenecks. In this work, we present a multi-axis MLP based architecture called MAXIM, that can serve as an efficient and flexible general-purpose vision backbone for image processing tasks. MAXIM uses a UNet-shaped hierarchical structure and supports long-range interactions enabled by spatially-gated MLPs. Specifically, MAXIM contains two MLP-based building blocks: a multi-axis gated MLP that allows for efficient and scalable spatial mixing of local and global visual cues, and a cross-gating block, an alternative to cross-attention, which accounts for cross-feature conditioning. Both these modules are exclusively based on MLPs, but also benefit from being both global and `fully-convolutional', two properties that are desirable for image processing. Our extensive experimental results show that the proposed MAXIM model achieves state-of-the-art performance on more than ten benchmarks across a range of image processing tasks, including denoising, deblurring, deraining, dehazing, and enhancement while requiring fewer or comparable numbers of parameters and FLOPs than competitive models. The source code and trained models will be available at \url{https://github.com/google-research/maxim}.

\end{abstract}


\section{Introduction}
\label{sec:intro}

Image processing tasks, such as restoration and enhancement, are important computer vision problems, which aim to produce a desired output from a degraded input.
Various types of degradations may require different image enhancement treatments, such as denoising, deblurring, super-resolution, dehazing, low-light enhancement, and so on.
Given the increased availability of curated large-scale training datasets, recent high-performing approaches~\cite{zamir2020learning,zamir2021multi,zhu2012deconvolving,yli_comisr_iccv2021,cho2021rethinking,chen2021pre,matlin2012removal,meng2020gia,chen2020proxiqa,liang2021swinir,delbracio2020projected} based on highly designed convolutional neural network (CNN) have demonstrated state-of-the-art (SOTA) performance on many tasks.

\begin{figure}[!t]
\center
\footnotesize
\setlength{\tabcolsep}{0pt}
\def\yheight{0.185}
\begin{tabular}{@{}c!{}c!{}c!{}c!{}c@{}}
\includegraphics[height=\yheight\textwidth]{ 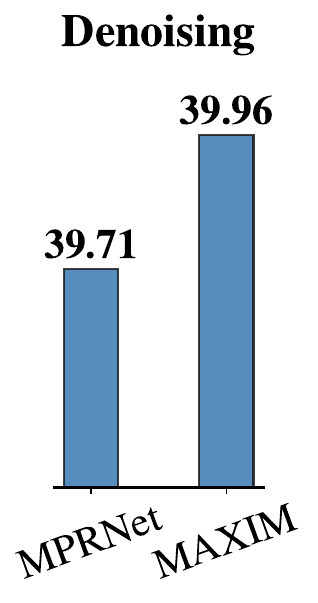}  &  
\includegraphics[height=\yheight\textwidth]{ 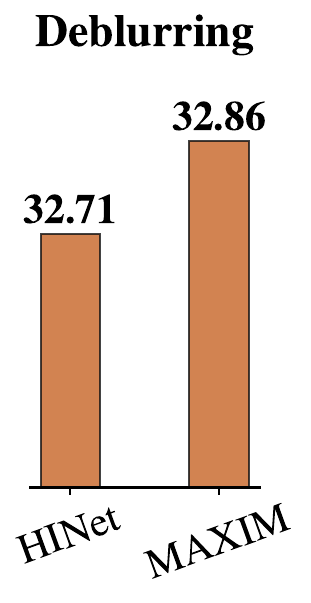} &
\includegraphics[height=\yheight\textwidth]{ 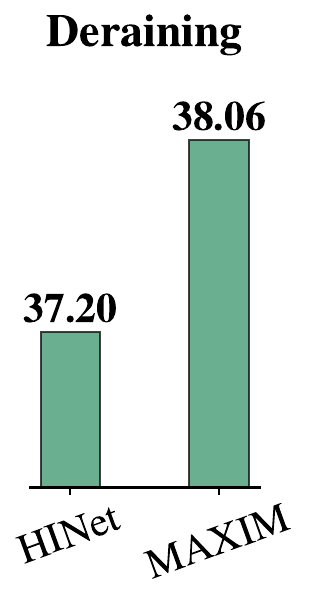} &
\includegraphics[height=\yheight\textwidth]{ 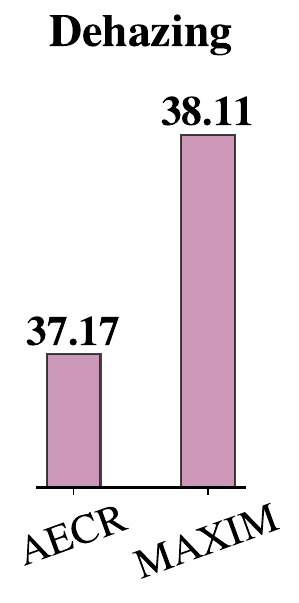} &
\includegraphics[height=\yheight\textwidth]{ 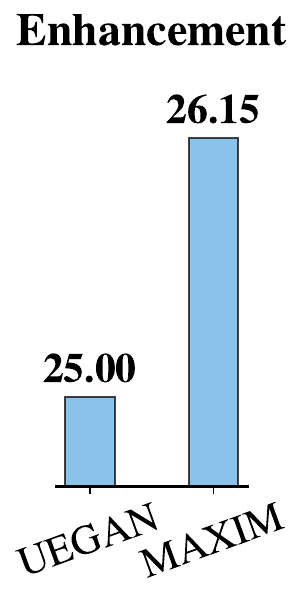} \\
\end{tabular}
\vspace{-3mm}
\caption{Our proposed MAXIM model significantly advances state-of-the-art performance on five image processing tasks in terms of PSNR: 1) Denoising ($+$\textbf{0.24} dB on SIDD~\cite{abdelhamed2018high}), 2) Deblurring ($+$\textbf{0.15} dB on GoPro~\cite{nah2017deep}) 3) Deraining ($+$\textbf{0.86} dB on Rain100L~\cite{yang2017deep}), 4) Dehazing ($+$\textbf{0.94} dB on RESIDE~\cite{li2019benchmarking}), and 5) Retouching (Enhancement) ($+$\textbf{1.15} dB on FiveK~\cite{bychkovsky2011learning}).}
\label{fig:teaser-figure}
\vspace{-5mm}
\end{figure}

Improving the architectural design of the underlying model is one of the keys to improving the performance of most computer vision tasks, including image restoration. 
Numerous researchers have invented or borrowed individual modules or building blocks and implemented them into low-level vision tasks, including residual learning~\cite{zhang2017beyond,ledig2017photo,wang2018esrgan}, dense connections~\cite{zhang2018residual,wang2018esrgan}, hierarchical structures~\cite{lai2017deep,kupyn2019deblurgan,jiang2021enlightengan}, multi-stage frameworks~\cite{zamir2021multi,chen2021hinet,jiang2020multi,zhang2019deep}, and attention mechanisms ~\cite{zamir2020learning,zamir2021multi,niu2020single,wang2019spatial}.

Recent research explorations on Vision Transformers (ViT)~\cite{dosovitskiy2021an,carion2020end,liu2021Swin} have exemplified their great potential as alternatives to the go-to CNN models.
The elegance of ViT~\cite{dosovitskiy2021an} has also motivated similar model designs with simpler global operators such as MLP-Mixer~\cite{tolstikhin2021mlp}, gMLP~\cite{liu2021pay}, GFNet~\cite{rao2021global}, and FNet~\cite{lee2021fnet}, to name a few.
Despite successful applications to many high-level tasks~\cite{dosovitskiy2021an,liu2021Swin,vaswani2021scaling,arnab2021vivit,xu2022v2x,xie2021segformer,sun2020transtrack}, the efficacy of these \textit{global} models on low-level enhancement and restoration problems has not been studied extensively.
The pioneering works on Transformers for low-level vision~\cite{chen2021pre,cao2021video} directly applied full self-attention, which only accepts relatively small patches of fixed sizes (e.g., 48$\times$48).
Such a strategy will inevitably cause patch boundary artifacts when applied on larger images using cropping~\cite{chen2021pre}.
Local-attention based Transformers~\cite{liang2021swinir,wang2021uformer} ameliorate this issue, but they are also constrained to have limited sizes of receptive field, or to lose non-locality~\cite{wang2018non,dosovitskiy2021an}, which is a compelling property of Transformers and MLP models relative to hierarchical CNNs.

To overcome these issues, we propose a generic image processing network, dubbed \textbf{{MAXIM}}, for low-level vision tasks.
A key design element of MAXIM is the use of \textit{multi-axis} approach (\cref{ssec:multi-axis-gating-mlp}) that captures both local and global interactions in parallel.
By mixing information on \textit{a single axis} for each branch, this MLP-based operator becomes `\textit{fully-convolutional}' and scales linearly with respect to image size, which significantly increases its flexibility for dense image processing tasks.
We also define and build a pure MLP-based cross-gating module, which adaptively \textit{gate} the skip-connections in the neck of MAXIM using the same multi-axis approach, and which further boosts performance.
Inspired by recent restoration models, we develop a simple but effective multi-stage, multi-scale architecture consisting of a stack of MAXIM backbones. MAXIM achieves strong performance on a range of image processing tasks, while requiring very few number of parameters and FLOPs.
Our contributions are:

\vspace{-2mm}
\begin{itemize}[leftmargin=*]
\itemsep0em 
    \item A novel and generic architecture for image processing, dubbed MAXIM, using a stack of encoder-decoder backbones, supervised by a multi-scale, multi-stage loss.
    \vspace{-0.6mm}
    \item A multi-axis gated MLP module tailored for low-level vision tasks, which always enjoys a global receptive field, with linear complexity relative to image size.
    \vspace{-0.6mm}
    \item A cross gating block that cross-conditions two separate features, which is also global and fully-convolutional.
    \vspace{-0.6mm}
    \item Extensive experiments show that MAXIM achieves SOTA results on more than 10 datasets including denoising, deblurring, deraining, dehazing, and enhancement.
\end{itemize}

\section{Related Work}
\label{sec:related-work}

\noindent\textbf{Restoration models.}
Driven by recent enormous efforts on building vision benchmarks, learning-based models, especially CNN models, have been developed that attain state-of-the-art performance on a wide variety of image enhancement tasks~\cite{zamir2021multi,chen2021hinet,jiang2021enlightengan,yli_comisr_iccv2021,shen2019human,chen2020proxiqa,chen2021pre,liang2021swinir}. These increased performance gains can be mainly attributed to novel architecture designs, and/or task-specific modules and units. For instance, UNet~\cite{ronneberger2015u} has incubated many successful encoder-decoder designs~\cite{zamir2021multi,jiang2021enlightengan,cho2021rethinking} for image restoration that improve on earlier single-scale feature processing models~\cite{zhang2017beyond,li2017aod}. Advanced components developed for high-level vision tasks have been brought into low-level vision tasks as well. Residual and dense connections~\cite{zhang2017beyond,ledig2017photo,wang2018esrgan, zhang2018residual,wang2018esrgan}, the multi-scale feature learning~\cite{kupyn2019deblurgan,wang2021uformer,cho2021rethinking}, attention mechanisms~\cite{zamir2020learning,zamir2021multi,niu2020single,wang2019spatial,zhang2018residual}, and non-local networks~\cite{wang2018non,liu2018non,zhang2018residual} are such good examples.
Recently, \textit{multi-stage} networks~\cite{zamir2021multi,chen2021hinet,jiang2020multi,zhang2019deep} have attained promising results relative to the aforementioned \textit{single-stage} models on the challenging deblurring and deraining tasks~\cite{dong2020multi,jiang2020multi,zamir2021multi}.
These multi-stage frameworks are generally inspired by their success on higher-level problems such as pose estimation~\cite{chen2018cascaded, li2019rethinking}, action segmentation~\cite{farha2019ms,li2020ms}, and image generation~\cite{zhang2017stackgan,zhang2018stackgan++}.

\noindent\textbf{Low-level vision Transformers.}
Transformers were originally proposed for NLP tasks~\cite{vaswani2017attention}, where multi-head self-attention and feed-forward MLP layers are stacked to capture non-local interactions between words.
Dosovitskiy \etal coined the term Vision Transformer (ViT)~\cite{dosovitskiy2021an}, and demonstrated the first pure Transformer model for image recognition. Several recent studies explored Transformers for low-level vision problems, \eg, the pioneering pre-trained image processing Transformer (IPT)~\cite{chen2021pre}.
Similar to ViT, IPT directly applies vanilla Transformers to image patches. The authors of~\cite{cao2021video} presented a spatial-temporal convolutional self-attention network that exploits local information for video super-resolution. More recently, Swin-IR~\cite{liang2021swinir} and UFormer~\cite{wang2021uformer} apply efficient window-based local attention models on a range of image restoration tasks. 

\noindent\textbf{MLP vision models.}
More recently, several authors have argued that when using a patch-based architecture as in ViT, the necessity of complex self-attention mechanisms becomes questionable. For instance, MLP-Mixer~\cite{tolstikhin2021mlp} adopts a simple token-mixing MLP to replace self-attention in ViT, resulting in an all-MLP architecture. The authors of~\cite{liu2021pay} proposed the gMLP, which applies a spatial gating unit on visual tokens. ResMLP~\cite{touvron2021resmlp} adopts an Affine transformation as a substitute to Layer Normalization for acceleration. Very recent techniques such as FNet~\cite{lee2021fnet} and GFNet~\cite{rao2021global} demonstrate the simple Fourier Transform can be used as a competitive alternative to either self-attention or MLPs. 



\begin{figure*}[!htb]
\centering
\includegraphics[width=0.95\linewidth]{ 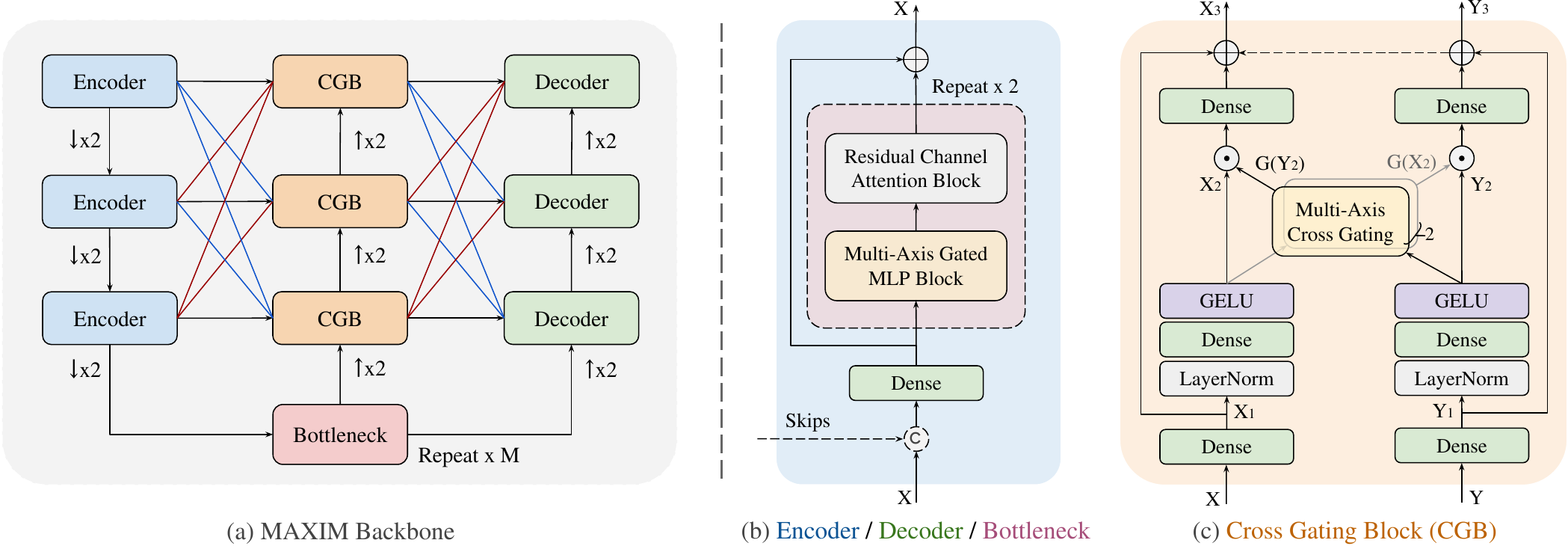}
\vspace{-2mm}
\caption{\textbf{MAXIM architecture.} We take (a) an encoder-decoder backbone with each (b) encoder, decoder, and bottleneck containing a multi-axis gated MLP block (\cref{fig:multi-axis-gated-mlp-block}) as well as a residual channel attention block. The model is further boosted by (c) a cross gating block which allows global contextual features to gate the skip-connections. More detailed description can be found in~\cref{sec:model-details}.}
\vspace{-3mm}
\label{fig:MAXIM-backbone}
\end{figure*}

\section{Our Approach: MAXIM}
\label{sec:approach}

We present, to the best of our knowledge, the first effective general-purpose MLP architecture for low-level vision, which we call \textbf{M}ulti-\textbf{AXI}s \textbf{M}LP for image processing (\textbf{MAXIM}). 
Unlike previous low-level Transformers~\cite{chen2021pre,cao2021video,liang2021swinir,wang2021uformer}, MAXIM has several desired properties, making it intriguing for image processing tasks.
First, MAXIM expresses global receptive fields on arbitrarily large images with linear complexity; Second, it directly supports arbitrary input resolutions, \ie, being fully-convolutional; Lastly, it provides a balanced design of local (\texttt{Conv}) and global (\texttt{MLP}) blocks, outperforming SOTA methods without the necessity for large-scale pre-training~\cite{chen2021pre}. 

\subsection{Main Backbone}
\label{ssec:main-architecture}

The MAXIM backbone (\cref{fig:MAXIM-backbone}a) follows the encoder-decoder design principles that originated with UNet~\cite{ronneberger2015u}.
We have observed that operators having small footprints such as \texttt{Conv3x3} are essential to the performance of UNet-like networks. Thus, we rely on a hybrid model design for each block (\cref{fig:MAXIM-backbone}b) -- \texttt{Conv} for local, and \texttt{MLP} for long-range interactions -- to make the most of them.


To allow long-range spatial mixing at different scales, we insert the multi-axis gated MLP block (MAB) into each encoder, decoder, and bottleneck (\cref{fig:MAXIM-backbone}b), with a residual channel attention block (RCAB)~\cite{woo2018cbam,zamir2021multi} (\texttt{LayerNorm}-\texttt{Conv}-\texttt{LeakyReLU}-\texttt{Conv}-\texttt{SE}~\cite{hu2018squeeze}) stacked subsequently.
Inspired by the gated filtering of skip connections~\cite{oktay2018attention, petit2021u}, we extend the gated MLP (gMLP) to build a cross gating block (CGB, \cref{fig:MAXIM-backbone}c), which is an efficient 2nd-order alternative to cross-attention (3rd-order correlations), to interact, or condition two distinct features.
We leverage the global features from \textcolor{brickred}{Bottleneck} (\cref{fig:MAXIM-backbone}a) to gate the skip connections, while propagating the refined global features upwards to the next CGB.
Multi-scale feature fusion~\cite{sun2019deep,zamir2020learning,cho2021rethinking} (\textcolor{brickred}{red} and \textcolor{blueish}{blue} lines) is utilized to aggregate multi-level information in the Encoder$\rightarrow$CGB and CGB$\rightarrow$Decoder dataflow.

\subsection{Multi-Axis Gated MLP}
\label{ssec:multi-axis-gating-mlp}

\begin{figure*}[!htb]
\centering
 \includegraphics[width=0.95\linewidth]{ 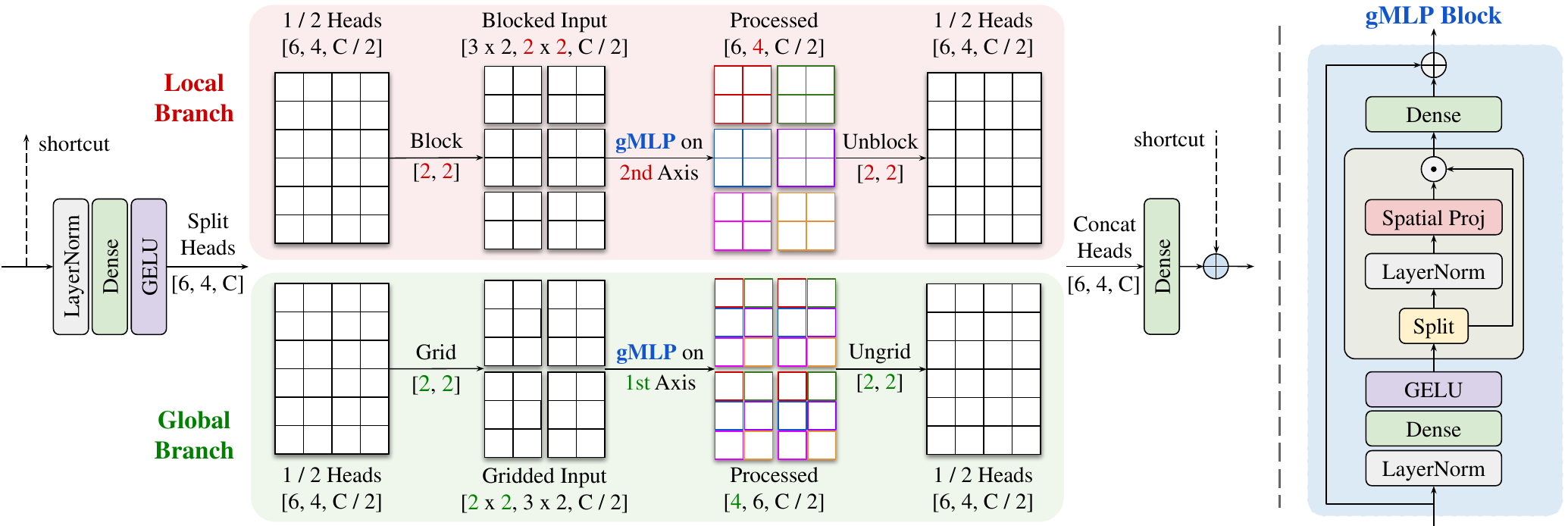}
 \caption{\textbf{Multi-axis gated MLP block} (best viewed in color). The input is first projected to a $[6,4,C]$ feature, then split into two heads. In the \textcolor{brickred}{local branch}, the half head is \textit{blocked} into $3\times 2$ non-overlapping $[\textcolor{brickred}{2},\textcolor{brickred}{2},C/2]$ patches, while we \textit{grid} the other half using a $\textcolor{dark_green}{2}\times \textcolor{dark_green}{2}$ grid in the \textcolor{dark_green}{global branch}. We only apply the gMLP block~\cite{liu2021pay} (illustrated in the right \textcolor{blueish}{gMLP Block}) on \textit{a single axis} of each branch - the \textcolor{brickred}{2nd} axis for the local branch and the \textcolor{dark_green}{1st} axis for the global branch, while shared along the other spatial dimensions. The gMLP operators, which run in parallel, correspond to local and global (dilated) attended regions, as illustrated with different colors (\ie, the same color are spatially mixed using the gMLP operator). Our proposed block expresses both global and local receptive fields on arbitrary input resolutions.}
 \label{fig:multi-axis-gated-mlp-block}
 \vspace{-3mm}
\end{figure*}

Our work is inspired by the multi-axis blocked self-attention proposed in~\cite{zhao2021improved}, which performs attention on more than a single axis.
The attentions performed on two axes on blocked images correspond to two forms of sparse self-attention, namely regional and dilated attention.
Despite capturing local and global information in parallel, this module cannot accommodate image restoration or enhancement tasks where the test images are often of arbitrary sizes.

We improve the `\textit{multi-axis}' concept for image processing tasks, by building a (split-head) multi-axis gated MLP block (MAB), as shown in \cref{fig:multi-axis-gated-mlp-block}.
Instead of applying multi-axis attention in a single layer~\cite{zhao2021improved}, we split in half the heads first, each being partitioned independently.
In the \textcolor{brickred}{local branch}, the half head of a feature of size $(H,W,C/2)$ is \textit{blocked} into a tensor of shape $(\frac{H}{b}\times\frac{W}{b},b\times b,C/2)$, representing partitioning into non-overlapping windows each with size of $(b\times b)$; in the \textcolor{dark_green}{global branch}, the other half head is \textit{gridded} into the shape $(d\times d,\frac{H}{{d}}\times\frac{W}{d},C/2)$ using a fixed $(d\times d)$ grid, with each window having size $(\frac{H}{{d}}\times\frac{W}{{d}})$.
For visualization, we set $b=\textcolor{brickred}{2},d=\textcolor{dark_green}{2}$ in \cref{fig:multi-axis-gated-mlp-block}. 
To make it \textit{fully-convolutional}, we only apply the gated MLP (gMLP) block~\cite{liu2021pay} on \textit{a single axis} of each branch -- the \textcolor{brickred}{$2$nd axis} for the local branch and the \textcolor{dark_green}{$1$st axis} for the global branch -- while sharing parameters on the other spatial axes.
Intuively, applying multi-axis gMLPs in parallel correspond to local and global (dilated) mixing of spatial information, respectively.
Finally, the processed heads are concatenated and projected to reduce the number of channels, which are further combined using the long skip-connection from the input.
It is worth noting that this approach provides an advantage for our model over methods that process fixed-size image patches~\cite{chen2021pre} by avoiding patch boundary artifacts.


\noindent\textbf{Complexity analysis.}
The computational complexity of our proposed Multi-Axis gMLP block (MAB) is:
\vspace{-1mm}
\begin{equation}
\label{eq:complexity}
\small
\Omega(\mathrm{MAB})=\underbrace{d^2HWC}_{\textcolor{dark_green}{\mathrm{Global\  gMLP}}}+\underbrace{b^2HWC}_{\textcolor{brickred}{\mathrm{Local\  gMLP}}}+\underbrace{10HWC^2}_{\textcolor{blueish}{\mathrm{Dense\ layers}}},
\end{equation}
which is \textit{linear} with respect to image size $HW$, while other global models like ViT, Mixer, and gMLP are \textit{quadratic}.

\noindent\textbf{Universality of the multi-axis approach.} Our proposed parallel multi-axis module (\cref{fig:multi-axis-gated-mlp-block}) presents a principled way to apply 1D operators on 2D images in a scalable manner.
It also allows for significant flexibility and universality.
For example, a straightforward replacement of a gMLP with a spatial MLP~\cite{tolstikhin2021mlp}, self-attention~\cite{dosovitskiy2021an}, or even Fourier Transform~\cite{rao2021global,lee2021fnet} leads to a family of MAXIM variants (see~\cref{ssec:ablation}D), all sharing globality and fully-convolutionality.
It is also easily extensible to \textit{any} future 1D operator that may be defined on, \eg, Language models.


\subsection{Cross Gating MLP Block}
\label{ssec:cross-gating-block-mlp}

A common improvement over UNet is to leverage contextual features to selectively \textit{gate} feature propagation in skip-connections~\cite{oktay2018attention,petit2021u}, which is often achieved by using cross-attention~\cite{vaswani2017attention,chen2021crossvit}.
Here we build an effective alternative, namely cross-gating block (CGB, \cref{fig:MAXIM-backbone}c), as an extension of MAB (\cref{ssec:multi-axis-gating-mlp}) which can only process a single feature.
CGB can be regarded as a more general conditioning layer that interacts with multiple features~\cite{vaswani2017attention,perez2018film,chen2021crossvit}.
We follow similar design patterns as those used in MAB.

To be more specific, let $\mathbf{X},\mathbf{Y}$ be two input features, and $\mathbf{X}_1,\mathbf{Y}_1\in\mathbb{R}^{H\times W\times C}$ be the features projected after the first \texttt{Dense} layers in \cref{fig:MAXIM-backbone}c. Input projections are then applied:
\begin{equation}
\label{eq:cross-gating-block-input-projection}
\small
\mathbf{X}_2=\sigma(\mathbf{W}_1 \mathsf{LN}(\mathbf{X}_{1}))\ ,\ \ \mathbf{Y}_2=\sigma(\mathbf{W}_2 \mathsf{LN}(\mathbf{Y}_{1}))
\end{equation}
where $\sigma$ is the $\mathsf{GELU}$ activation~\cite{hendrycks2016gaussian}, $\mathsf{LN}$ is Layer Normalization~\cite{ba2016layer}, and $\mathbf{W}_1,\mathbf{W}_2$ are \texttt{MLP} projection matrices.
The multi-axis blocked gating weights are computed from $\mathbf{X}_2,\mathbf{Y}_2$, respectively, but applied \textit{reciprocally}:
\begin{equation}
\label{eq:cross-gating-block-cross-gating}
\small
\hat{\mathbf{X}}=\mathbf{X}_2\odot \mathrm{G}(\mathbf{Y}_2)\ ,\ \  \hat{\mathbf{Y}}=\mathbf{Y}_2\odot \mathrm{G}(\mathbf{X}_2)
\end{equation}
where $\odot$ represents element-wise multiplication, and the function $\mathrm{G}(\cdot)$ extracts multi-axis cross gating weights from the input using our proposed multi-axis approach (\cref{ssec:multi-axis-gating-mlp}):
\begin{equation}
\label{eq:cross-gating-block-gating-function}
\small
\mathrm{G}(\mathbf{x})=\mathbf{W}_5([\mathbf{W}_3\mathsf{Block}_{b}(\mathbf{z_1}), \mathbf{W}_4\mathsf{Grid}_{d}(\mathbf{z_2})]) 
\end{equation}
where $[\cdot,\cdot]$ denotes concatenation.
Here $(\mathbf{z_1},\mathbf{z_2})$ are two independent heads split from $\mathbf{z}$ along the channel dimension, where $\mathbf{z}$ represents the projected features $\mathbf{x}$ after activation:
\begin{equation}
\small
\label{eq:cross-gating-block-gating-function-input-proj}
[\mathbf{z_1},\mathbf{z_2}]=\mathbf{z}=\sigma(\mathbf{W}_6 \mathsf{LN}(\mathbf{x})),
\end{equation}
and $\mathbf{W}_3,\mathbf{W}_4$ are spatial projection matrices applied on the \textcolor{brickred}{2nd} and \textcolor{dark_green}{1st} axis of the blocked/gridded features having fixed window size $b\times b$ ($\mathsf{Block}_{b}$), and fixed grid size of $d\times d$ ($\mathsf{Grid}_{d}$), respectively.
Finally, we adopt residual connection from the inputs, following an output channel-projection that maintains the same channel dimensions as the inputs ($\mathbf{X}_1,\mathbf{Y}_1$), using projection matrices $\mathbf{W}_7$, $\mathbf{W}_8$, denoted by
\begin{equation}
\label{eq:cross-gating-block-output-proj}
\small
\mathbf{X}_{3}=\mathbf{X}_{1}+\mathbf{W}_7\hat{\mathbf{X}}\ ,\ \  \mathbf{Y}_{3}=\mathbf{Y}_{1}+\mathbf{W}_8\hat{\mathbf{Y}}.
\end{equation}
The complexity of CGB is also tightly-bounded by \cref{eq:complexity}.

\subsection{Multi-Stage Multi-Scale Framework}
\label{ssec:multi-stage-multi-scale-framework}

We further adopt a multi-stage framework because we find it more effective, as compared to scaling up the model width or height (see ablation \cref{ssec:ablation}A).
We deem full resolution processing~\cite{park2020multi,ren2019progressive,chen2021hinet} a better approach than a multi-patch hierarchy~\cite{suin2020spatially,zamir2021multi,zhang2019deep}, since the latter would potentially induce boundary effects across patches. 
To impose stronger supervision, we apply a multi-scale approach~\cite{li2019rethinking,chen2018cascaded,cho2021rethinking} at each stage to help the network learn. We leverage the supervised attention module~\cite{zamir2021multi} to propagate attentive features progressively along the stages.
We leverage the cross-gating block (\cref{ssec:cross-gating-block-mlp}) for cross-stage feature fusion. We refer the reader to~\cref{fig:multi-stage-framework} for details.

\begin{table}[!t]
\centering
\scriptsize
\setlength{\tabcolsep}{3pt}
\renewcommand{\arraystretch}{1.1}
\begin{tabular}{l|cc|cc||cc}
& \multicolumn{2}{c}{SIDD~\cite{abdelhamed2018high}} & \multicolumn{2}{c}{DND~\cite{plotz2017benchmarking}} & \multicolumn{2}{c}{Average} \\ 
Method & PSNR$\uparrow$ & SSIM$\uparrow$ & PSNR$\uparrow$ & SSIM$\uparrow$ &  PSNR$\uparrow$ & SSIM$\uparrow$ \\
\toprule
DnCNN~\cite{zhang2017beyond} & 23.66 & 0.583 & 32.43 & 0.790 & 28.04 & 0.686 \\
MLP~\cite{burger2012image} & 24.71 & 0.641 & 34.23 & 0.833 & 29.47 & 0.737 \\
BM3D~\cite{dabov2007image} & 35.65 & 0.685 & 34.51 & 0.851 & 35.08 & 0.768 \\
CBDNet\textcolor{red}{*}~\cite{guo2019toward} & 30.78 & 0.801 & 38.06 & 0.942 & 34.42 & 0.872  \\
RIDNet\textcolor{red}{*}~\cite{anwar2019real} & 38.71 & 0.951 & 39.26 & 0.953 & 38.99 & 0.952 \\
AINDNet\textcolor{red}{*}~\cite{kim2020transfer} & 38.95 & 0.952 & 39.37 & 0.951 & 39.16 & 0.952 \\
VDN~\cite{yue2019variational} & 39.28 & 0.956 & 39.38 & 0.952  & 39.33 & 0.954 \\
SADNet\textcolor{red}{*}~\cite{chang2020spatial} & 39.46 & 0.957 & 39.59 & 0.952 & 39.53 & 0.955 \\
CycleISP\textcolor{red}{*}~\cite{zamir2020cycleisp} & 39.52 & 0.957 & 39.56 & \textbf{0.956} & 39.54 & \underline{0.957} \\
MIRNet~\cite{zamir2020learning} & \underline{39.72} & \underline{0.959} & \textbf{39.88} & \textbf{0.956} & \underline{39.80} & \textbf{0.958} \\
MPRNet~\cite{zamir2021multi} & {39.71} & {0.958} & {39.80} & \underline{0.954} & 39.76 & 0.956 \\
\midrule
MAXIM-3S & \textbf{39.96} & \textbf{0.960} & \underline{39.84} & \underline{0.954} & \textbf{39.90} & \underline{0.957} \\
\bottomrule
\end{tabular}
\caption{Denoising results. Our model is only trained on SIDD~\cite{abdelhamed2018high} and evaluated on SIDD~\cite{abdelhamed2018high} and DND~\cite{plotz2017benchmarking}, where \textcolor{red}{*} denotes methods using additional training data.}
\label{tab:denoising}
\end{table}

Formally, given an input image $\mathbf{I}\in \mathbb{R}^{H\times W\times3}$, we first extract its multi-scale variants by downscaling: $\mathbf{I}_n,\ n=1,...,N$. MAXIM predicts multi-scale restored outputs at each stage $s$ of $S$ stages, yielding a total of $S\times N$ outputs: $\mathbf{R}_{s,n}$. Despite being multi-stage, MAXIM is trained \textit{end-to-end} with losses accumulating across stages and scales:
\begin{equation}
\label{eq:loss}
\small
\mathcal{L}=\sum_{s=1}^S\sum_{n=1}^N [\mathcal{L}_{char}(\mathbf{R}_{s,n}, \mathbf{T}_{n})+\lambda\mathcal{L}_{freq}(\mathbf{R}_{s,n}, \mathbf{T}_{n})],
\end{equation}
where $\mathbf{T}_n$ denotes (bilinearly-rescaled) multi-scale target images, and $\mathcal{L}_{char}$ is the Charbonnier loss~\cite{zamir2021multi}:
\begin{equation}
\label{eq:char-loss}
\small
\mathcal{L}_{char}(\mathbf{R},\mathbf{T})=\sqrt{\|\mathbf{R}-\mathbf{T}\|^2+\epsilon^2},
\end{equation}
where we set $\epsilon=10^{-3}$. $\mathcal{L}_{freq}$ is the frequency reconstruction loss that enforces high-frequency details~\cite{jiang2021focal,cho2021rethinking}:
\begin{equation}
\label{eq:freq-loss}
\small
\mathcal{L}_{freq}(\mathbf{R},\mathbf{T})=\|\mathcal{F}(\mathbf{R})-\mathcal{F}(\mathbf{T})\|_1
\end{equation}
where $\mathcal{F}(\cdot)$ represents the 2D Fast Fourier Transform. We used $\lambda=0.1$ as the weighting factor in all experiments.

\section{Experiments}
\label{sec:experiments}

We aim at building a generic backbone for a broad spectrum of image processing tasks. 
Thus, we evaluated MAXIM on five different tasks: (1) denoising, (2) deblurring, (3) deraining, (4) dehazing, and (5) enhancement (retouching) on \textbf{17} different datasets (summarized in~\cref{tab:datasets}.
More comprehensive results and visualizations can be found in~\cref{sec:more-visual-comparisons}.

\subsection{Experimental Setup}
\label{ssec:experiment-setup}

\noindent\textbf{Datasets and metrics.} We measured PSNR and SSIM~\cite{wang2004image} metrics between ground truth and predicted images to make quantitative comparisons. We used SIDD~\cite{abdelhamed2018high} and DND~\cite{plotz2017benchmarking} for denoising, GoPro~\cite{nah2017deep}, HIDE~\cite{shen2019human}, and RealBlur~\cite{rim2020real} for debluring, a combined dataset Rain13k used in~\cite{zamir2021multi} for deraining. The RESIDE~\cite{li2019benchmarking} is used for dehazing, while Five-K\cite{bychkovsky2011learning} and LOL~\cite{wei2018deep} are evaluated for enhancement.

\begin{figure*}[!ht]
\centering
\footnotesize
\def\yem{1pt}
\def\xwidth{0.99}
\def\xxxwidth{0.12\textwidth}
\setlength{\tabcolsep}{1pt}
\begin{tabular}{C{\xxxwidth}C{\xxxwidth}C{\xxxwidth}C{\xxxwidth}C{\xxxwidth}C{\xxxwidth}C{\xxxwidth}C{\xxxwidth}}
\multicolumn{8}{c}{
\includegraphics[width=\xwidth\linewidth]{ 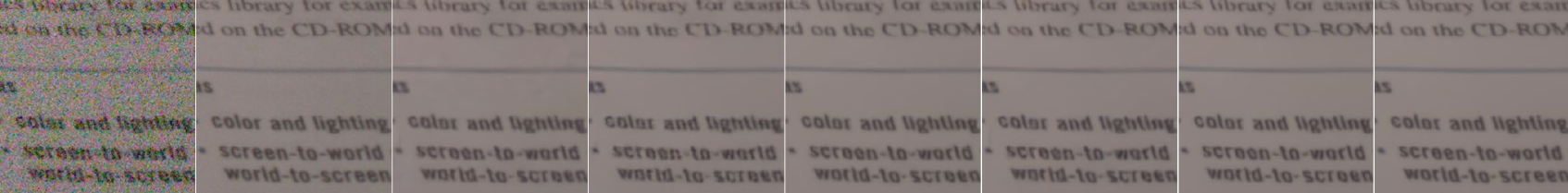}
} \\
Input & Target & VDN~\cite{yue2019variational} & DANet~\cite{yue2020dual} & MIRNet~\cite{zamir2020learning} & CycleISP~\cite{zamir2020cycleisp} & MPRNet~\cite{zamir2021multi} & MAXIM-3S \\
\end{tabular}
\caption{Denoising comparisons. The example from SIDD~\cite{abdelhamed2018high} shows that our method produces cleaner denoising results.}
\label{fig:denoising-visuals}
\end{figure*}

\begin{table}[!t]
\centering
\scriptsize
\setlength{\tabcolsep}{1.6pt}
\renewcommand{\arraystretch}{1.1}
\begin{tabular}{l|cc|cc||cc}
& \multicolumn{2}{c}{GoPro~\cite{nah2017deep}} & \multicolumn{2}{c}{HIDE~\cite{shen2019human}} & \multicolumn{2}{c}{Average} \\ 
Method & PSNR$\uparrow$ & SSIM$\uparrow$ & PSNR$\uparrow$ & SSIM$\uparrow$ & PSNR$\uparrow$ & SSIM$\uparrow$ \\
\toprule
DeblurGAN~\cite{kupyn2018deblurgan} & 28.70 & 0.858 & 24.51 & 0.871 & 26.61 & 0.865 \\
Nah \etal~\cite{nah2017deep} & 29.08 & 0.914 & 25.73 & 0.874  & 27.41 & 0.894 \\
Zhang \etal~\cite{zhang2018dynamic} & 29.19 & 0.931 & - & -  & - & - \\
DeblurGAN-v2~\cite{kupyn2019deblurgan} & 29.55 & 0.934 & 26.61 & 0.875 & 28.08  & 0.905 \\
SRN~\cite{tao2018scale} & 30.26 & 0.934 & 28.36 & 0.915  & 29.31 &  0.925 \\
Shen \etal~\cite{shen2019human} & - & - & 28.89 & 0.930 & - & - \\
Gao \etal~\cite{gao2019dynamic} & 30.90 & 0.935 & 29.11 & 0.913 &  30.01 & 0.924 \\
DBGAN~\cite{zhang2020deblurring} & 31.10 & 0.942 & 28.94 & 0.915 &  30.02 & 0.929 \\
MT-RNN~\cite{park2020multi} & 31.15 & 0.945 & 29.15 & 0.918 & 30.15 & 0.932\\
DMPHN~\cite{zhang2019deep} & 31.20 & 0.940 & 29.09 & 0.924  & 30.15 & 0.932 \\
Suin \etal~\cite{suin2020spatially} & 31.85 & 0.948 & 29.98 & 0.930 & 30.92 & 0.939 \\
MPRNet~\cite{zamir2021multi} & {32.66} & \underline{0.959} & \underline{30.96} & \underline{0.939} & \underline{31.81} & \underline{0.949} \\
Pretrained-IPT~\cite{chen2021pre} & 32.58 & - & - & -  & - & - \\
MIMO-UNet+~\cite{cho2021rethinking} & 32.45 & 0.957 & 29.99 & 0.930 & 31.22 & 0.944 \\
HINet~\cite{chen2021hinet} & \underline{32.71} & \underline{0.959} & {30.32} & {0.932} & 31.52 &  0.946 \\
\midrule
MAXIM-3S & \textbf{32.86} & \textbf{0.961} & \textbf{32.83} & \textbf{0.956} & \textbf{32.85} & \textbf{0.959} \\
\bottomrule
\end{tabular}
\caption{Deblurring results. Our model is trained on GoPro~\cite{nah2017deep} and evaluated on the GoPro and the HIDE dataset~\cite{shen2019human}.}
\label{tab:deblurring-gopro-hide}
\end{table}

\begin{table}[!t]
\centering
\scriptsize
\setlength{\tabcolsep}{1.6pt}
\renewcommand{\arraystretch}{1.1}
\begin{tabular}{l|cc|cc||cc}
& \multicolumn{2}{c}{RealBlur-R~\cite{rim2020real}} & \multicolumn{2}{c}{RealBlur-J~\cite{rim2020real}} & \multicolumn{2}{c}{Average} \\ 
Method & PSNR$\uparrow$ & SSIM$\uparrow$ & PSNR$\uparrow$ & SSIM$\uparrow$ & PSNR$\uparrow$ & SSIM$\uparrow$\\
\toprule
Hu \etal~\cite{hu2014deblurring} & 33.67 & 0.916 & 26.41 & 0.803 & 30.04  & 0.860 \\
Nah \etal~\cite{nah2017deep} & 32.51 & 0.841 & 27.87 & 0.827 & 30.19 & 0.834 \\
DeblurGAN~\cite{kupyn2018deblurgan} & 33.79 & 0.903 & 27.97 & 0.834 & 30.88 & 0.869 \\
Pan \etal~\cite{pan2016blind} & 34.01 & 0.916 & 27.22 & 0.790 & 30.62 & 0.853 \\
Xu \etal~\cite{xu2013unnatural} & 34.46 & 0.937 & 27.14 & 0.830 & 30.8 & 0.884 \\
DeblurGAN-v2~\cite{kupyn2019deblurgan} & 35.26 & 0.944 & 28.70 & 0.866 & 31.98 & 0.905 \\
Zhang \etal~\cite{zhang2018dynamic} & 35.48 & 0.947 & 27.80 & 0.847 & 31.64  & 0.897\\
SRN~\cite{tao2018scale} & 35.66 & 0.947 & 28.56 & 0.867 & 32.11 & 0.907 \\
DMPHN~\cite{zhang2019deep} &  35.70 & \underline{0.948} & 28.42 & 0.860 & 32.06 & 0.904 \\
MPRNet~\cite{zamir2021multi} & \textbf{35.99} & \textbf{0.952} & \underline{28.70} & \underline{0.873} & \textbf{32.35} & \textbf{0.913} \\
\hline
MAXIM-3S & \underline{35.78} & 0.947 & \textbf{28.83} & \textbf{0.875} & \underline{32.31} &  \underline{0.911} \\
\midrule[0.05pt]\midrule
\textcolor{red}{$^\dagger$}DeblurGAN-v2 & 36.44 & 0.935 & 29.69 & 0.870 & 33.07 & 0.903 \\
\textcolor{red}{$^\dagger$}SRN~\cite{tao2018scale} & 38.65 & \underline{0.965} & 31.38 & 0.909 & 35.02 & 0.937 \\
\textcolor{red}{$^\dagger$}MPRNet~\cite{zamir2021multi} & \underline{39.31} & \textbf{0.972} & 31.76 & \underline{0.922} & \underline{35.54} & \underline{0.947} \\
\textcolor{red}{$^\dagger$}MIMO-UNet+~\cite{cho2021rethinking} & - & - & \underline{32.05} & 0.921 & - & - \\
\hline
\textcolor{red}{$^\dagger$}MAXIM-3S & \textbf{39.45} & 0.962 & \textbf{32.84} & \textbf{0.935} & \textbf{36.15} & \textbf{0.949} \\
\bottomrule
\end{tabular}
\caption{Deblurring results on RealBlur~\cite{rim2020real}. \textcolor{red}{$^\dagger$} denotes methods that are trained on RealBlur, while those without \textcolor{red}{$^\dagger$} indicate methods trained only on GoPro.}
\label{tab:deblurring-realblur}
\end{table}

\begin{figure*}[!t]
\centering
\footnotesize
\def\yem{-2.5pt}
\def\xwidth{0.99}
\def\xxxwidth{0.12\textwidth}
\setlength{\tabcolsep}{1pt}
\begin{tabular}{C{\xxxwidth}C{\xxxwidth}C{\xxxwidth}C{\xxxwidth}C{\xxxwidth}C{\xxxwidth}C{\xxxwidth}C{\xxxwidth}}
\multicolumn{8}{c}{
\includegraphics[width=\xwidth\linewidth]{ 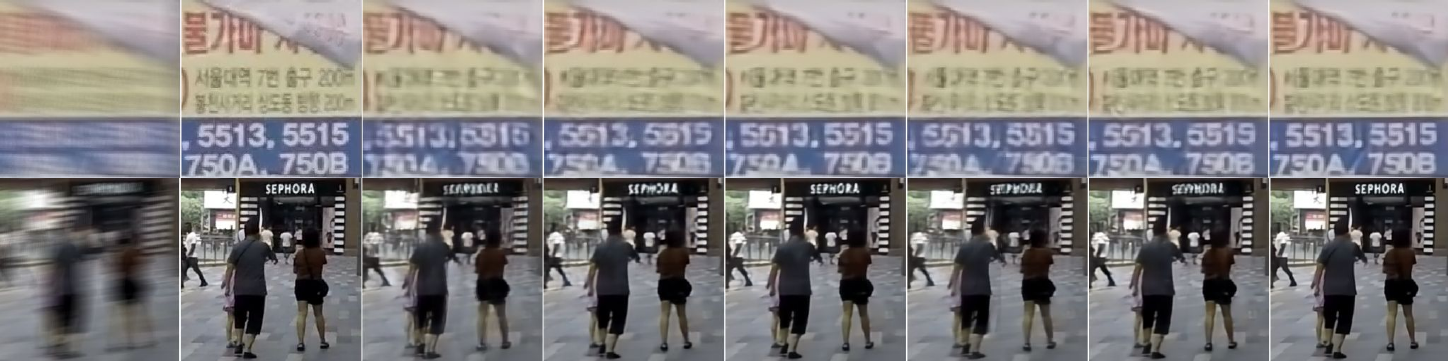}
} \\
Input & Target & DMPHN~\cite{zhang2019deep} & Suin \etal~\cite{suin2020spatially} & MPRNet~\cite{zamir2021multi} & HINet~\cite{chen2021hinet} & MIMO-UNet~\cite{cho2021rethinking} & MAXIM-3S
\end{tabular}
\caption{Deblurring comparisons. The top row shows an example from GoPro~\cite{nah2017deep} while the second row shows one from HIDE~\cite{shen2019human}.}
\label{fig:deblurring-visuals}
\vspace{-3mm}
\end{figure*}

\noindent\textbf{Training details.} Our proposed MAXIM model is end-to-end trainable and requires neither large-scale pretraining nor progressive training. The network is trained on $256\!\times\!256$ random-cropped patches. We train different iterations for each task. We used random horizontal and vertical flips, $90^{\circ}$ rotation, and MixUp~\cite{zhang2017mixup} with probability $0.5$ for data augmentation. We used the Adam optimizer~\cite{kingma2014adam} with an initial learning rate of $2\!\times\!10^{-4}$, which are steadily decreased to $10^{-7}$ with the cosine annealing decay~\cite{loshchilov2016sgdr}. When testing, we padded the input images to be a multiplier of $64\!\times\!64$ using symmetric padding on both sides. After inference, we cropped the padded image back to original size. 
More training details on each task can be found in~\cref{sec:datasets-n-training-details}.

\noindent\textbf{Architectural configuration.}
We designed two MAXIM variants: a two-stage model called MAXIM-\textcolor{r}{2S}, and a three-stage model, MAXIM-\textcolor{b}{3S}, for different tasks. We start with $32$ initial channels for feature extraction, with 3 downsampling layers, where the features contract from $256^2\times 32$, $128^2\times64$, $64^2\times 128$, to $32^2\times 256$ processed by \textit{two} \textcolor{brickred}{Bottlenecks} (\cref{fig:MAXIM-backbone}a), then symmetrically expanded back to full resolution. The number of parameters and required FLOPs of MAXIM-\textcolor{r}{2S} and MAXIM-\textcolor{b}{3S}, when applied on a $256\times 256$ image are shown in the last two rows of \cref{tab:ablation}A.

\subsection{Main Results}
\label{ssec:main-results}

\noindent\textbf{Denoising.}
We report in \cref{tab:denoising} numerical comparisons on the SIDD~\cite{abdelhamed2018high} and DND~\cite{plotz2017benchmarking} datasets. As may be seen, our method outperformed previous SOTA techniques, \eg, MIRNet~\cite{zamir2020learning} by \textbf{0.24 dB} of PSNR on SIDD while obtaining competitive PSNR (39.84 dB) on DND. \cref{fig:denoising-visuals} shows visual results on SIDD. Our method clearly removes real noise while maintaining fine details, yielding visually pleasant results to the other methods.

\noindent\textbf{Deblurring.} \cref{tab:deblurring-gopro-hide} shows the quantitative comparison of MAXIM-\textcolor{b}{3S} against SOTA deblurring methods on two synthetic blur datasets: GoPro~\cite{nah2017deep} and HIDE~\cite{shen2019human}. Our method achieves \textbf{0.15} dB gain in PSNR over the previous best model HINet~\cite{chen2021hinet}. It is notable that the GoPro-trained MAXIM-\textcolor{b}{3S} model generalizes extremely well on the HIDE dataset, setting new SOTA PSNR values: \textbf{32.83} dB. We also evaluated on real-world blurry images from RealBlur~\cite{rim2020real} under two settings: (1) directly applied the GoPro-trained model on RealBlur, and (2) fine-tuned the model on RealBlur. Under setting (1), MAXIM-\textcolor{b}{3S} ranked \textit{first} on RealBlur-J subset while obtaining the top two performance on RealBlur-R. \cref{fig:deblurring-visuals} shows visual comparisons of the evaluated models on GoPro~\cite{nah2017deep}, HIDE~\cite{shen2019human} and RealBlur~\cite{rim2020real}, respectively. It may be observed that our model recovers \textbf{text} extremely well, which may be attributed to the use of multi-axis MLP module within each block that globally aggregates repeated patterns across various scales.

\begin{figure*}[!t]
\centering
\footnotesize
\def\yem{-2.5pt}
\def\xwidth{0.99}
\def\xxxwidth{0.12\textwidth}
\setlength{\tabcolsep}{1pt}
\begin{tabular}{C{\xxxwidth}C{\xxxwidth}C{\xxxwidth}C{\xxxwidth}C{\xxxwidth}C{\xxxwidth}C{\xxxwidth}C{\xxxwidth}}
\multicolumn{8}{c}{
\includegraphics[width=\xwidth\linewidth]{ 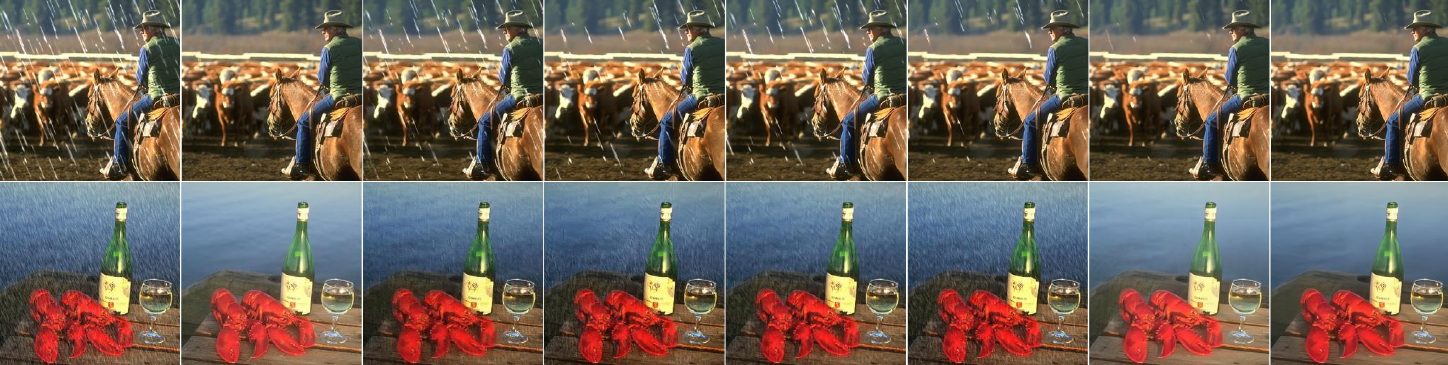}
} \\
Input & Target & RESCAN~\cite{li2018recurrent} & PreNet~\cite{ren2019progressive} & MSPFN~\cite{jiang2020multi} & MPRNet~\cite{zamir2021multi} & HINet~\cite{chen2021hinet} & MAXIM-2S \\
\end{tabular}
\caption{Deraining comparisons. The top and bottom rows present examples from Rain100L~\cite{yang2017deep} and Test100~\cite{zhang2019image}, respectively, demonstrating the ability of MAXIM to remove rain streaks while recovering more details, hence yielding more visually pleasant results.}
\label{fig:deraining-visuals}
\end{figure*}

\begin{figure*}[!t]
\centering
\footnotesize
\def\yem{-2.5pt}
\def\xwidth{0.99}
\def\xxxwidth{0.12\textwidth}
\setlength{\tabcolsep}{1pt}
\begin{tabular}{C{\xxxwidth}C{\xxxwidth}C{\xxxwidth}C{\xxxwidth}C{\xxxwidth}C{\xxxwidth}C{\xxxwidth}C{\xxxwidth}}
\multicolumn{8}{c}{
\includegraphics[width=\xwidth\linewidth]{ 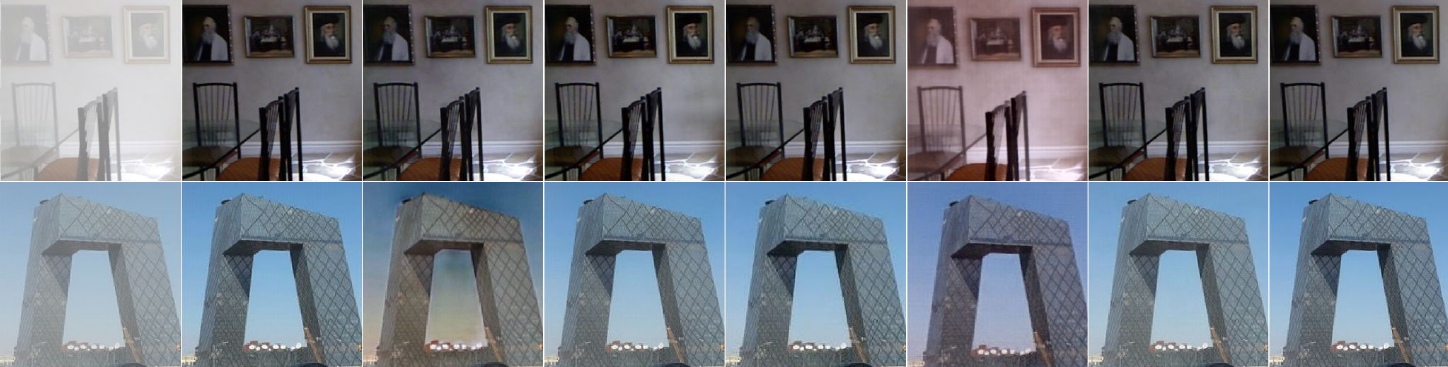}
} \\
Input & Target & GCANet~\cite{chen2019gated} & GridDehaze~\cite{liu2019griddehazenet} & DuRN~\cite{liu2019dual} & MSBDN~\cite{dong2020multi} & FFA-Net~\cite{qin2020ffa} & MAXIM-2S \\
\end{tabular}
\caption{Dehazing comparisons. The top and bottom rows exemplify visual results from the SOTS indoor and outdoor sets~\cite{li2019benchmarking}.}
\label{fig:dehazing-visuals}
\end{figure*}

\begin{table*}[!t]
\scriptsize
\centering
\setlength{\tabcolsep}{1.25pt}
\renewcommand{\arraystretch}{1.1}
\begin{tabular}{@{}lr@{}}
\begin{minipage}[t]{.68\linewidth}
\begin{tabular}{@{}l|cc|cc|cc|cc|cc||cc@{}}
& \multicolumn{2}{c}{Rain100L~\cite{yang2017deep}} & \multicolumn{2}{c}{Rain100H~\cite{yang2017deep}} & 
\multicolumn{2}{c}{Test100~\cite{zhang2019image}} & 
\multicolumn{2}{c}{Test1200~\cite{zhang2018density}} & 
\multicolumn{2}{c}{Test2800~\cite{fu2017removing}} & \multicolumn{2}{c}{Average}  \\ 
Method & PSNR$\uparrow$ & SSIM$\uparrow$ & PSNR$\uparrow$ & SSIM$\uparrow$  & PSNR$\uparrow$ & SSIM$\uparrow$ & PSNR$\uparrow$ & SSIM$\uparrow$  & PSNR$\uparrow$ & SSIM$\uparrow$  & PSNR$\uparrow$ & SSIM$\uparrow$ \\
\toprule
DerainNet~\cite{fu2017clearing} & 27.03 & 0.884 & 14.92 & 0.592 & 22.77 & 0.810 & 23.38 & 0.835 & 24.31 & 0.861 & 22.48 & 0.796  \\
SEMI~\cite{wei2019semi} & 25.03 & 0.842 & 16.56 & 0.486 & 22.35 & 0.788 & 26.05 & 0.822 & 24.43 & 0.782 & 22.88 & 0.744 \\
DIDMDN~\cite{zhang2018density} & 25.23 & 0.741 & 17.35 & 0.524 & 22.56 & 0.818 & 29.65 & 0.901 & 28.13 & 0.867 & 24.58 & 0.770 \\
UMRL~\cite{yasarla2019uncertainty} & 29.18 & 0.923 & 26.01 & 0.832 & 24.41 & 0.829 & 30.55 & 0.910 & 29.97 & 0.905 &   28.02 &  0.880\\ 
RESCAN~\cite{li2018recurrent} & 29.80 & 0.881 & 26.36 & 0.786 & 25.00 & 0.835 & 30.51 & 0.882 & 31.29 & 0.904 & 28.59 & 0.857 \\
PreNet~\cite{ren2019progressive} & 32.44 & 0.950 & 26.77 & 0.858 & 24.81 & 0.851 & 31.36 & 0.911 & 31.75 & 0.916 & 29.42 & 0.897 \\
MSPFN~\cite{jiang2020multi} & 32.40 & 0.933 & 28.66 & 0.860 & 27.50 & 0.876 & 32.39 & 0.916 & 32.82 & 0.930 & 30.75 & 0.903 \\
MPRNet~\cite{zamir2021multi} & 36.40 & 0.965 & 30.41 & 0.890 & \underline{30.27} & 0.897 & \underline{32.91} & 0.916 & 33.64 & 0.938 & 32.73 & 0.921 \\
HINet~\cite{chen2021hinet} & \underline{37.20} & \underline{0.969} & \underline{30.63} & \underline{0.893} & 30.26 & \underline{0.905} & \textbf{33.01} & \underline{0.918} & \textbf{33.87} & \underline{0.940} & \underline{33.00} & \underline{0.925} \\
\midrule
MAXIM-2S & \textbf{38.06} &	\textbf{0.977} & \textbf{30.81} & \textbf{0.903} & \textbf{31.17} & \textbf{0.922}	& \underline{32.37} & \textbf{0.922}	& \underline{33.80} & \textbf{0.943} & \textbf{33.24} & \textbf{0.933} \\
\bottomrule
\end{tabular}
\caption{Deraining comparisons. Our method consistently yields better quality metrics with respect to both PSNR or SSIM on all the tested datasets: Rain100L~\cite{yang2017deep}, Rain100H~\cite{yang2017deep}, Test100~\cite{zhang2019image}, Test1200~\cite{zhang2018density}, Test2800~\cite{fu2017removing}}
\label{tab:deraining}
\end{minipage} &

\hspace{6pt}

\begin{minipage}[t]{.29\linewidth}
\begin{tabular}{@{}l|cc|cc@{}}
& \multicolumn{2}{c}{SOTS-Indoor} & \multicolumn{2}{c}{SOTS-Outdoor} \\ 
Method & PSNR$\uparrow$ & SSIM$\uparrow$ & PSNR$\uparrow$ & SSIM$\uparrow$ \\
\toprule
DehazeNet~\cite{cai2016dehazenet} & 21.14 & 0.847 & 22.46 & 0.851 \\
GFN~\cite{Ren-CVPR-2018} & 22.30 & 0 880 & 21.55 & 0.844 \\
GCANet~\cite{chen2019gated} & 30.23 & 0.959 & 19.98  & 0.704 \\
GridDehaze~\cite{liu2019griddehazenet} & 32.14 & 0.983 & 30.86 & 0.981 \\
GMAN~\cite{liu2019single} & 27.93 & 0.896 & 28.47 & 0.944 \\
MSBDN~\cite{dong2020multi} & 33.79 & 0.984 & 23.36 & 0.875 \\
DuRN~\cite{liu2019dual} & 32.12 & 0.980 & 24.47 & 0.839 \\
FFA-Net~\cite{qin2020ffa} & 36.39 & 0.989 & \underline{33.57} & \underline{0.984} \\
AECR-Net~\cite{wu2021contrastive} & \underline{37.17} & \underline{0.990} & - & - \\
\midrule
MAXIM-2S & \textbf{38.11} & \textbf{0.991} & \textbf{34.19} & \textbf{0.985}  \\
\bottomrule
\end{tabular}
\caption{Dehazing comparisons. Our model achieved the best results on both indoor and outdoor scenes.}
\label{tab:dehazing}
\end{minipage} \\
\end{tabular}
\vspace{-3mm}
\end{table*}

\begin{figure*}[!t]
\centering
\footnotesize
\def\yem{-2.5pt}
\def\xwidth{0.99}
\def\xxxwidth{0.12\textwidth}
\setlength{\tabcolsep}{1pt}
\begin{tabular}{C{\xxxwidth}C{\xxxwidth}C{\xxxwidth}C{\xxxwidth}C{\xxxwidth}C{\xxxwidth}C{\xxxwidth}C{\xxxwidth}}
\multicolumn{8}{c}{
\includegraphics[width=\xwidth\linewidth]{ 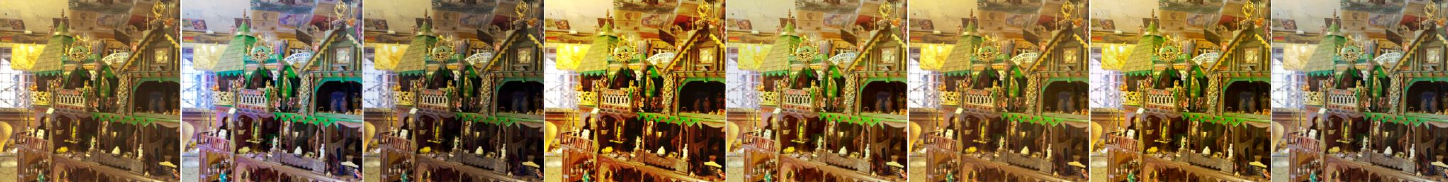}
} \\
Input & Target & CycleGAN~\cite{zhu2017unpaired} & Exposure~\cite{hu2018exposure} & DPE~\cite{chen2018deep} & EnlightenGAN & UEGAN~\cite{ni2020towards} & MAXIM-2S \\
\multicolumn{8}{c}{
\includegraphics[width=\xwidth\linewidth]{ 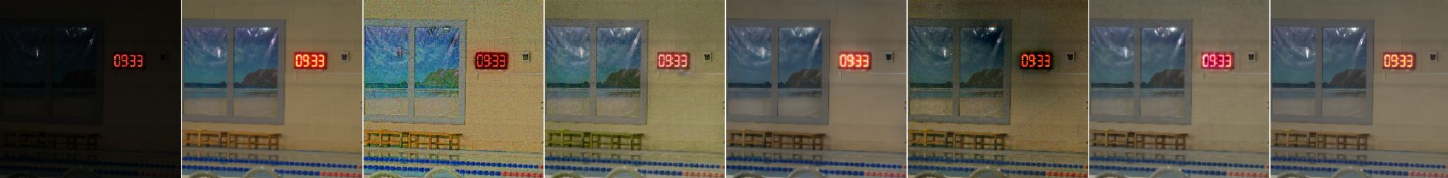}
} \\
Input & Target & Retinex~\cite{wei2018deep} & GLAD~\cite{wang2018gladnet} & KinD~\cite{zhang2019kindling} & EnlightenGAN & MIRNet~\cite{zamir2020learning} & MAXIM-2S \\
\end{tabular}
\caption{Retouching and low-light enhancement comparisons. The top row shows an example from the MIT-Adobe FiveK dataset~\cite{bychkovsky2011learning}, while the bottom row exemplifies a comparison from LOL~\cite{wei2018deep}. Our model generated variegated and more naturalistic colors (top) for retouching, while achieving clearer and brighter visual enhancements in the bottom example.}
\label{fig:enhancement-visuals}
\vspace{-3mm}
\end{figure*}


\noindent\textbf{Deraining.} Following previous work~\cite{jiang2020multi,zamir2021multi}, we computed the performance metrics using the Y channel (in YCbCr color space). \cref{tab:deraining} shows quantitative comparisons with previous methods. As may be seen, our model improved over the SOTA performances on all datasets. The average PSNR gain of our model over the previous best model HINet~\cite{chen2021hinet} is \textbf{0.24} dB. We demonstrate some challenging examples in \cref{fig:deraining-visuals}, which demonstrates that our method consistently delivered faithfully recovered images without introducing any noticeable visual  artifacts. 

\noindent\textbf{Dehazing.} We report our comparisons against SOTA models in \cref{tab:dehazing}. Our model surpassed the previous best model by \textbf{0.94} dB and \textbf{0.62} dB of PSNR on the SOTS~\cite{li2019benchmarking} indoor and outdoor sets. \cref{fig:dehazing-visuals} shows that our model recovered images of better quality on both flat regions as well as textures, while achieving a harmonious global tone.

\noindent\textbf{Enhancement / Retouching.} As \cref{tab:enhancement} illustrates, our model achieved the best PSNR and SSIM values on FiveK~\cite{bychkovsky2011learning} and LOL~\cite{wei2018deep}, respectively. As the top row of \cref{fig:enhancement-visuals} suggests, MAXIM recovered diverse naturalistic colors as compared to other techniques. Regarding the bottom example, while MIRNet~\cite{zamir2020learning} obtained a higher PSNR, we consistently observed that our model attains visually better quality with sharper details and less noise. Moreover, the far more perceptually relevant SSIM index indicates a significant advantage of MAXIM-\textcolor{r}{2S} relative to MIRNet.

\noindent\textbf{Other benchmarks.} Due to space limitations, we detail the outcomes of our experiments on the REDS deblurring~\cite{Nah_2021_CVPR} and the Raindrop removal task~\cite{qian2018attentive} in~\cref{sec:additional-experiments}.

\subsection{Ablation}
\label{ssec:ablation}

We conduct extensive ablation studies to validate the proposed multi-axis gated MLP block, cross-gating block, and multi-stage multi-scale architecture. The evaluations were performed on the GoPro dataset~\cite{nah2017deep} trained on image patches of size $256\times 256$ for $10^6$ iterations. We used the MAXIM-\textcolor{r}{2S} model as the test-bed for Ablation-A and -B.

\noindent\textbf{A. Individual components.}
We conducted an ablation by progressively adding (1) inter-stage cross-gating blocks (CGB\textsubscript{IS}), (2) a supervised attention module (SAM), (3) cross-stage cross-gating blocks (CGB\textsubscript{CS}, and (4) the multi-scale supervision (MS-Sp). \cref{tab:ablation}A indicates a PSNR gain of 0.25, 0.63, 0.36, 0.26 dB for each respective component.

\begin{table}[!h]
\centering
\scriptsize
\setlength{\tabcolsep}{4pt}
\renewcommand{\arraystretch}{1.1}
\begin{tabular}{l|cc||l|cc}
& \multicolumn{2}{c}{FiveK~\cite{bychkovsky2011learning}} & & \multicolumn{2}{c}{LOL~\cite{wei2018deep}} \\ 
Method & PSNR$\uparrow$ & SSIM$\uparrow$ & Method & PSNR$\uparrow$ & SSIM$\uparrow$ \\
\toprule
CycleGAN~\cite{zhu2017unpaired} & 18.23 & 0.835 &  Retinex~\cite{wei2018deep} & 16.77 & 0.559 \\
Exposure~\cite{hu2018exposure} & 22.35 & 0.861 & GLAD~\cite{wang2018gladnet} & 19.71 & 0.703 \\
EnlightenGAN & 17.74 & 0.828 & EnlightenGAN & 17.48 & 0.657 \\
DPE~\cite{chen2018deep} & 24.08 & 0.922 & KinD~\cite{zhang2019kindling} & 20.37 & 0.804 \\
UEGAN~\cite{ni2020towards} & \underline{25.00} & \underline{0.929} & MIRNet~\cite{zamir2020learning} & \textbf{24.14} & \underline{0.830} \\
\midrule
MAXIM-2S & \textbf{26.15} & \textbf{0.945} & MAXIM-2S & \underline{23.43} & \textbf{0.863}    \\
\bottomrule
\end{tabular}
\caption{Enhancement results on FiveK~\cite{bychkovsky2011learning} and LOL~\cite{wei2018deep}.}
\label{tab:enhancement}
\end{table}

\noindent\textbf{B. Effects of multi-axis approach.}
We further examined the necessity of our proposed multi-axis approach, as shown in \cref{tab:ablation}B. We conducted experiments over (1) baseline UNet, (2) by adding the local branch of MAB (MAB\textsubscript{$\ell$}), (3) by adding the global branch of MAB (MAB\textsubscript{$g$}), (4) by adding the local branch of CGB (CGB\textsubscript{$\ell$}), (5) by adding the global branch of CGB (CGB\textsubscript{$g$}). Note that the huge jump (+1.04 dB) of PSNR by adding MAB\textsubscript{$\ell$} can be largely attributed to the addition of input and output channel projection layers, because we also observe a high performance of \textbf{31.42} dB PSNR if only MAB\textsubscript{$g$} is added. Overall, we observed a \textit{major} improvement when including MAB, and a relatively \textit{minor} gain when adding CGB.

\noindent\textbf{C. Why multi-stage?} Towards understanding this, we scaled up MAXIM in terms of width (channels), depth (downscaling steps), and the number of stages. \cref{tab:ablation}C suggests that packing the backbone into multi-stages yields the best performance \vs complexity tradeoff (32.44 dB, 22.2 M, 339.2 G), compared to making it wider or deeper.

\noindent\textbf{D. Beyond gMLP: the MAXIM families.}
As described in \cref{ssec:multi-axis-gating-mlp}, our proposed multi-axis approach (\cref{fig:multi-axis-gated-mlp-block}) offers a scalable way of applying \textit{any} 1D operators on (high-resolution) images, with linear complexity relative to image size while maintaining fully-convolutional. We conducted a pilot study using MAXIM-\textcolor{g}{1S} and -\textcolor{r}{2S} on SIDD~\cite{abdelhamed2018high} to explore the MAXIM families: MAXIM-FFT, -MLP, -gMLP (modeled in this paper), -SA, where we use the Fourier Transform filter~\cite{rao2021global,lee2021fnet}, spatial MLP~\cite{tolstikhin2021mlp}, gMLP~\cite{liu2021pay}, and self-attention~\cite{dosovitskiy2021an} on spatial axes using the same multi-axis approach (\cref{fig:multi-axis-gated-mlp-block}). As \cref{tab:ablation}D shows, the gMLP and self-attention variants achieved the best performance, while the FFT and MLP families were more computationally efficient. We leave deeper explorations to future works.




\begin{table}[!t]
\centering
\scriptsize
\setlength{\tabcolsep}{1.2pt}
\renewcommand{\arraystretch}{1.1}
\begin{tabular}[t]{@{}c@{}}

\scriptsize
\setlength{\tabcolsep}{1.8pt}
\begin{tabular}{cccc|c}
 \scriptsize{CGB\textsubscript{IS}}  &  \scriptsize{SAM} &  \scriptsize{CGB\textsubscript{CS}} &  \scriptsize{MS-Sp} &  \scriptsize{PSNR}   \\
\midrule
  &   &   &  & 30.73 \\ \arrayrulecolor{lightgray}\hline
 \cmark &   &   & & 30.98   \\ \arrayrulecolor{lightgray}\hline
 \cmark & \cmark &   &  & 31.61   \\ \arrayrulecolor{lightgray}\hline
 \cmark & \cmark & \cmark & & 31.97  \\ \arrayrulecolor{lightgray}\hline 
 \cmark & \cmark & \cmark & \cmark & 32.23 \\ 
\arrayrulecolor{black}\bottomrule
\multicolumn{5}{c}{{A. Individual components.}}\\
\end{tabular}

\hfill
\scriptsize
\setlength{\tabcolsep}{1.8pt}
\begin{tabular}{cccc|c}
\scriptsize{MAB\textsubscript{$\ell$}} & \scriptsize{MAB\textsubscript{$g$}} & \scriptsize{CGB\textsubscript{$\ell$}} & \scriptsize{CGB\textsubscript{$g$}} & \scriptsize{PSNR}    \\
\midrule
  &   &   &   & 30.48   \\ \arrayrulecolor{lightgray}\hline
 \cmark &   &   &  & 31.52    \\ \arrayrulecolor{lightgray}\hline
 \cmark & \cmark &   &  & 31.68 \\ \arrayrulecolor{lightgray}\hline
 \cmark & \cmark & \cmark & & 31.84 \\ \arrayrulecolor{lightgray}\hline
 \cmark & \cmark & \cmark & \cmark & 31.91 \\
\arrayrulecolor{black}\bottomrule
\multicolumn{5}{c}{{B. Effects of multi-axis approach.}}\\

\end{tabular} 
\\\\

\def\arraystretch{1.26}
\setlength{\tabcolsep}{1.8pt}
\scriptsize
\begin{tabular}{l|ccc|c|r|r}
 & S & W & D & PSNR & Params & FLOPs \\
\midrule
 Base & 1 & 32 & 3 & 31.08 & 6.1M & 93.6G  \\ 
 \arrayrulecolor{lightgray}\hline
 \multirow{2}{*}{Wider} & 1 & 64 & 3 & 32.09 & 19.4M & 309.9G \\
                        & 1 & 96 & 3 & 32.31 & 41.7M & 648.9G \\
 \arrayrulecolor{lightgray}\hline
 \multirow{2}{*}{Deeper} & 1 & 32 & 4 & 31.17 & 19.8M & 121.6G  \\
& 1 & 32 & 5 & 31.43 & 75.0M & 153.4G \\ \arrayrulecolor{lightgray}\hline

 More          & 2 & 32 & 3 & 31.82 & 14.1M & 216.4G \\
 stages    & 3 & 32 & 3 & 32.44 & 22.2M & 339.2G \\
\arrayrulecolor{black}\bottomrule
\multicolumn{7}{c}{{C. Why multi-stage?}}\\
\end{tabular}
\hfill

\scriptsize
\setlength{\tabcolsep}{1.8pt}
\def\arraystretch{1.12}
\begin{tabular}{l|r|r|r}
Variant & PSNR & Params & FLOPs \\
\midrule

M1-FFT &  39.67 & 4.1M & 71G \\ 
M1-MLP & 39.75 & 5.4M & 83G \\ 
M1-gMLP & 39.80 & 6.1M & 93G \\ 
M1-SA & 39.79 & 5.3M & 111G \\ \arrayrulecolor{lightgray}\hline
M2-FFT &  39.74 & 10.1M & 172G \\ 
M2-MLP & 39.70 & 12.7M & 195G \\ 
M2-gMLP & 39.83 & 14.1M & 216G \\ 
M2-SA & 39.85 & 12.5M & 250G \\

\arrayrulecolor{black}\bottomrule
\multicolumn{4}{c}{{D. Beyond gMLP.}}\\
\end{tabular} 
\\

\end{tabular}
\caption{Ablation studies. Components in subtable A and B are defined in \cref{ssec:ablation}. S, W, and D denote the number of stages, width, and depth, respectively. M1 and M2 in subtable D denote MAXIM-1S and MAXIM-2S models, respectively.}
\label{tab:ablation}
\vspace{-1mm}
\end{table}

\section{Conclusion}
\label{sec:Conclusion}

We have presented a generic network for restoration or enhancement tasks, dubbed MAXIM, inspired by recently popular MLP-based global models. Our work suggests an effective and efficient approach for applying gMLP to low-level vision tasks to gain global attention, a missing attribute of basic CNNs. Our gMLP initialization of the MAXIM family significantly advances state-of-the-arts in several image enhancement and restoration tasks with moderate complexity. We demonstrate a few applications, but there are many more possibilities beyond the scope of this work which could significantly benefit by using MAXIM.
Our future work includes exploring more efficient models for extremely high-resolution image processing, as well as training large models that can adapt on multiple tasks.

\noindent\textbf{Broader impacts.} 
The proposed model can be used as an effective tool to enhance and retouch daily photos. 
However, enhancing techniques such as denoising and deblurring are vulnerable to malicious use for privacy concerns.
The models trained on specific data may express bias.
These issues should be responsibly taken care of by researchers.

\section{Acknowledgment}
We thank Junjie Ke, Mauricio Delbracio, Sungjoon Choi, Irene Zhu, Innfarn Yoo, Huiwen Chang, and Ce Liu for valuable discussions and feedback.


\appendix
\section{Appendix}
\label{sec:appendix}

\subsection{Datasets and Training Details}
\label{sec:datasets-n-training-details}

All the datasets used in the paper are summarized in \cref{tab:datasets}. We describe details of training for each dataset in the following. Note that we used the $\ell_2$ loss for the dehazing task while using the loss defined in the main paper for all the other tasks.

\noindent\textbf{Image Denoising.} We trained our model on $320$ high-resolution images provided in SIDD~\cite{abdelhamed2018high} and evaluated on 1,280 ($256\times 256$) and 1,000 ($512\times 512$) images provided by authors of SIDD~\cite{abdelhamed2018high} and DND~\cite{plotz2017benchmarking}, respectively. The results on DND were obtained via the online server~\cite{dnd-website}. 
We cropped the training images into $512\times 512$ patches with a stride of 256 to prepare the training patches. We trained the MAXIM-3S model for 600k steps with a batch size of 256.

\noindent\textbf{Image Deblurring.} We trained our model on 2,103 image pairs from GoPro~\cite{nah2017deep}. 
To demonstrate generalization ability, we evaluated our GoPro trained model on 1,111 pairs of the GoPro evaluation set, 2,025 images in the HIDE dataset~\cite{shen2019human}, as well as the RealBlur dataset~\cite{rim2020real}, which contains 980 paired images of camera JPEG output and RAW images, respectively. We cropped training images from GoPro into $512\times 512$ patches with a stride of 128 to generate training patches. 
We trained our MAXIM-3S model over 600k steps with a batch size of 256. For evaluation on RealBlur setting (2) (see main paper), we loaded the GoPro pre-trained checkpoint and fine-tuned for 70k and 15k iterations on RealBlur-J and RealBlur-R, respectively. Additionally, we trained our model on 24,000 images from the REDS dataset of the NTIRE 2021 Image Deblurring Challenge Track 2 JPEG artifacts~\cite{Nah_2021_CVPR}. 
For evaluation, we followed the settings in the NTIRE 2021 Challenge on Image Deblurring~\cite{nah2021ntire}, \ie, we used 300 images in the validation set of REDS. 
We trained from scratch for 10k epochs on REDS~\cite{Nah_2021_CVPR}.

\begin{table}[!t]
\centering
\scriptsize
\setlength{\tabcolsep}{4pt}
\renewcommand{\arraystretch}{1.1}
\begin{tabular}{l|l|rr|l}
 \textbf{Task}  & \textbf{Dataset}  & \textbf{\#Train} & \textbf{\#Test} & \textbf{Test Dubname} \\
 \toprule
 \multirow{2}{*}{\footnotesize\textbf{Denoising}} & SIDD~\cite{abdelhamed2018high} & 320 & 40 & SIDD \\
 & DND~\cite{plotz2017benchmarking} & 0 & 50 & DND \\
 \hline
 \multirow{5}{*}{\footnotesize\textbf{Deblurring}} & 
 GoPro~\cite{nah2017deep} & 2103 & 1111 & GoPro \\
 & HIDE~\cite{shen2019human} & 0 & 2025 & HIDE \\
 & RealBlur-J~\cite{rim2020real} & 3758 & 980 & RealBlur-J \\
 & RealBlur-R~\cite{rim2020real} & 3758 & 980 & RealBlur-R \\
 & REDS~\cite{Nah_2021_CVPR} & 24000 & 300 & REDS \\
 \hline
 \multirow{7}{*}{\footnotesize\textbf{Deraining}} & 
 Rain14000~\cite{fu2017removing} & 11200 & 2800 & Test2800 \\
 & Rain1800~\cite{yang2017deep} & 1800 & 0 & - \\
 & Rain800~\cite{zhang2019image} & 700 & 98 & Test100 \\
 & Rain100H~\cite{yang2017deep}  & 0 & 100 & Rain100H \\
 & Rain100L~\cite{yang2017deep}  & 0 & 100 & Rain100L  \\
 & Rain1200~\cite{zhang2018density} & 0  & 1200 & Test1200 \\
 & Rain12~\cite{li2016rain} & 12 & 0 & - \\ 
 & Raindrop~\cite{qian2018attentive} & 861 & 58 & Raindrop-A \\
 & Raindrop~\cite{qian2018attentive} & 0 & 239 & Raindrop-B \\
 \hline
 \multirow{2}{*}{\footnotesize\textbf{Dehazing}} 
 & RESIDE-ITS~\cite{li2019benchmarking} & 13990 & 500 & SOTS-Indoor \\
 & RESIDE-OTS~\cite{li2019benchmarking} & 313950 & 500 & SOTS-Outdoor \\
 \hline
{\footnotesize\textbf{Enhancement}} 
 & MIT-Adobe FiveK~\cite{bychkovsky2011learning} & 4500 & 500 & FiveK \\
{\footnotesize\textbf{(Retouching)}}  & LOL~\cite{wei2018deep} & 485 & 15 & LOL \\
\bottomrule
\end{tabular}
\caption{Dataset summary on five image processing tasks.}
\label{tab:datasets}
\end{table}

\begin{figure*}[!t]
 \centering
 \includegraphics[width=0.98\linewidth]{ 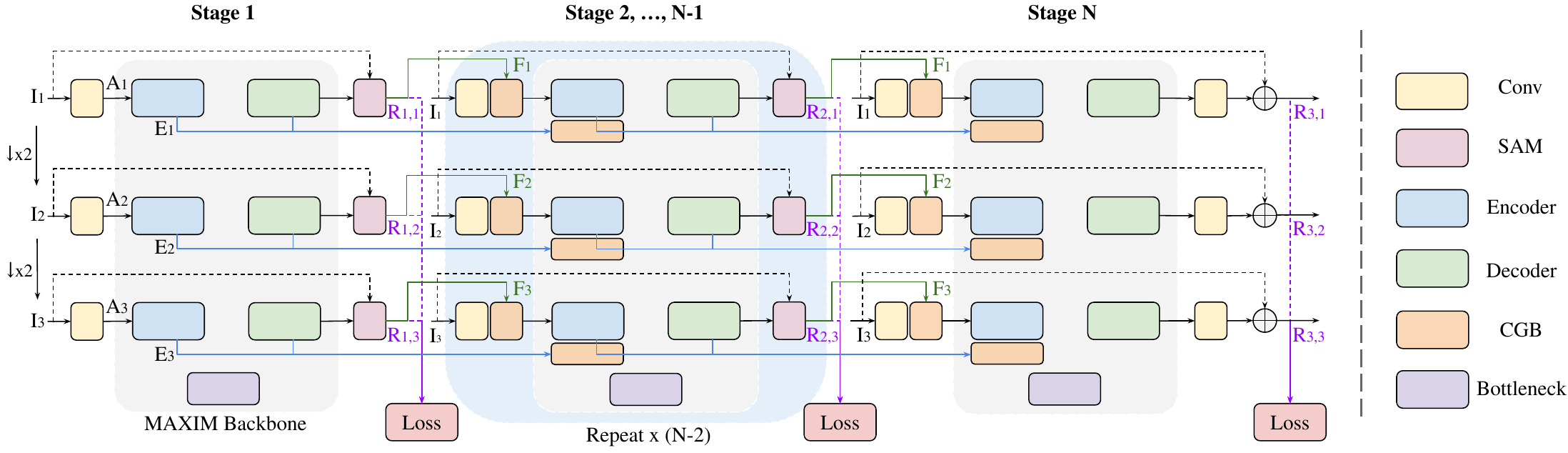}
 \caption{We adopt a general multi-stage framework to improve the performance of MAXIM for challenging restoration tasks. Inspired by~\cite{zamir2021multi, chen2021hinet}, we employ the supervised attention module (SAM) and cross-stage feature fusion to help later stages learn. Unlike previous approaches, our MAXIM backbone attains global perception at each layer in each stage due to the proposed multi-axis MLP approaches, making it more powerful in learning global interactions in both low-level and high-level features.}
 \label{fig:multi-stage-framework}
\end{figure*}

\noindent\textbf{Image Deraining.} Following~\cite{zamir2021multi, jiang2020multi}, we used a composite training set containing 13,712 clean-rain image pairs collected from multiple datasets~\cite{fu2017removing,yang2017deep,zhang2019image,yang2017deep,zhang2018density,li2016rain}. 
Evaluation was performed on five test sets, Rain100H~\cite{yang2017deep}, Rain100L~\cite{yang2017deep}, Test100~\cite{zhang2019image}, Test1200~\cite{zhang2018density}, and Test2800~\cite{fu2017removing}. 
We trained our MAXIM-2S model over 500k steps with a batch size of 512. For the raindrop removal task, we trained MAXIM-2S on 861 pairs of training images in Raindrop dataset~\cite{qian2018attentive} for 80k steps with a batch size of 512, and evaluate on testset A (58 images) and testset B (239 images), respectively.

\noindent\textbf{Image Dehazing.} The RESIDE dataset~\cite{li2019benchmarking} contains two subsets: Indoor Training Set (ITS) which contains 13,990 hazy images generated from 1399 clean ones, and Outdoor Training Set (OTS) that consists of 313,950 hazing images synthesized from 8,970 haze-free outdoor scenes. 
We evaluated our model on the Synthetic Objective Testing Set (SOTS)~\cite{li2019benchmarking}: 500 indoor images for ITS-trained, and 500 outdoor images for OTS-trained models, respectively. 
We trained for 10k and 500 epochs on RESIDE-ITS and RESIDE-OTS using the $\ell_2$ loss.

\noindent\textbf{Image Enhancement.} We used the MIT-Adobe FiveK~\cite{bychkovsky2011learning} dataset provided by~\cite{ni2020towards} for the retouching evaluation: the first 4,500 images for training and the rest 500 for testing. We cropped training images into $512\times 512$ patches with a stride of 256. We also used the LOL dataset~\cite{wei2018deep} which includes 500 pairs of images for low-light enhancement. We trained our model on 485 training images and evaluated on 15 test images. We trained for 14k and 180k steps on FiveK and LOL, respectively.

\subsection{Architecture Details}
\label{sec:model-details}

Our proposed general multi-stage and multi-scale framework is illustrated in \cref{fig:multi-stage-framework}, where each stage uses a single-stage MAXIM backbone, which is illustrated in the main paper. We leveraged the multi-scale input-output approach~\cite{cho2021rethinking} to deeply supervise each stage. Specifically, given an input image $\mathbf{I}\in\mathbb{R}^{H\times W\times 3}$, we used the nearest neighbour downscaling method~\cite{cho2021rethinking} to generate multi-scale input variants: $\mathbf{I}_n,\ n=1,2,3$, while we adopted a bilinear downscaler to produce the ground truth variants: $\mathbf{T}_n,\ n=1,2,3$. For each stage, we extracted shallow features from the inputs at each scale using \texttt{Conv3x3}. Except for the first stage, we fused the shallow features with attention features coming from the previous supervised attention module (SAM)~\cite{zamir2021multi} using a cross gating block (\textcolor{myyellow}{CGB}). We also employed cross-stage feature fusion~\cite{zamir2021multi,chen2021hinet} to help later stages, where the intermediate \textcolor{brandeisblue}{Encoder} and \textcolor{aogreen}{Decoder} features from the previous stage are fused with features encoded at the current stage using a \textcolor{myyellow}{CGB} (\textcolor{brandeisblue}{blue lines} in \cref{fig:multi-stage-framework}).

\subsubsection{Configurations}
\label{ssec:configurations}

The detailed specifications of the \textcolor{brandeisblue}{Encoder} part for a single-stage MAXIM are shown in \cref{tab:configurations}. We also provide the input and output shapes of each block and layer. Here Conv3x3\_s1\_w32 means a \texttt{Conv} layer with 3x3 kernels, stride 1, and 32 channels. MAB and RCAB are the two major components in \textcolor{brandeisblue}{Encoder} / \textcolor{aogreen}{Decoder} / \textcolor{brickred}{Bottleneck}. Note that in \textcolor{brickred}{Bottleneck} blocks, we use (\texttt{Conv1x1}) layers to replace \texttt{Conv3x3} in RCAB. 

\begin{table}[!t]
\centering
\scriptsize
\setlength{\tabcolsep}{4pt}
\renewcommand{\arraystretch}{1.1}
\begin{tabular}{c|l|l|c}
Depth  & Input shape & Output Shape & Layers  \\
\toprule
1  & $256^2\times 3$ &  $256^2\times 32$ & Conv3x3\_s1\_w32 \\
1 & $256^2\times 32$ & $256^2\times 32$ & CGB\textcolor{red}{*} ($b=d=16$) \\
1  & $256^2\times 32$ &  $256^2\times 32$ & Conv1x1\_s1\_w32 \\
\multirow{1}{*}{1} & \multirow{1}{*}{$256^2\times 32$} & \multirow{1}{*}{$256^2\times 32$} & \Big\{ $\begin{array}{@{}c@{}} \mathrm{MAB} (b=d=16)\\ \mathrm{RCAB} (3 \times 3,\  r=4) \end{array} \Big\} \times 2$ \\
1 & $256^2\times 32$ & $128^2\times 32$ & Conv3x3\_s2\_w32 \\
\midrule
2  & $128^2\times 32$ &  $128^2\times 64$ & Conv3x3\_s1\_w64 \\
2 & $128^2\times 64$ & $128^2\times 64$ & CGB\textcolor{red}{*} ($b=d=16$) \\
2  & $128^2\times 64$ &  $128^2\times 64$ & Conv1x1\_s1\_w64 \\
\multirow{1}{*}{2} & \multirow{1}{*}{$128^2\times 64$} & \multirow{1}{*}{$128^2\times 64$} & \Big\{ $\begin{array}{@{}c@{}} \mathrm{MAB} (b=d=16)\\ \mathrm{RCAB} (3 \times 3,\  r=4) \end{array} \Big\} \times 2$ \\
2 & $128^2\times 64$ & $64^2\times 64$ & Conv3x3\_s2\_w64 \\

\midrule
3  & $64^2\times 64$ &  $64^2\times 128$ & Conv3x3\_s1\_w128 \\
3  & $64^2\times 128$ &  $64^2\times 128$ & CGB\textcolor{red}{*} ($b=d=8$) \\
3  & $64^2\times 128$ &  $64^2\times 128$ & Conv1x1\_s1\_w128 \\
\multirow{1}{*}{3} & \multirow{1}{*}{$64^2\times 128$} & \multirow{1}{*}{$64^2\times 128$ } & \Big\{ $\begin{array}{@{}c@{}} \mathrm{MAB} (b=d=8)\\ \mathrm{RCAB} (3 \times 3,\  r=4) \end{array} \Big\} \times 2$ \\
3 & $64^2\times 128$ & $32^2\times 128$ & Conv3x3\_s2\_w128 \\

\midrule
4  & $32^2\times 128$ &  $32^2\times 256$ & Conv1x1\_s1\_w256 \\
\multirow{1}{*}{4} & \multirow{1}{*}{$32^2\times 256$} & \multirow{1}{*}{$32^2\times 256$} & \Big\{ $\begin{array}{@{}c@{}} \mathrm{MAB} (b=d=8)\\ \mathrm{RCAB} (1 \times 1,\  r=4) \end{array} \Big\} \times 2$ \\
\midrule
4 & $32^2\times 256$ & $32^2\times 256$ & Conv1x1\_s1\_w256 \\
\multirow{1}{*}{4} & \multirow{1}{*}{$32^2\times 256$} & \multirow{1}{*}{$32^2\times 256$} & \Big\{ $\begin{array}{@{}c@{}} \mathrm{MAB} (b=d=16)\\ \mathrm{RCAB} (1 \times 1,\  r=4) \end{array} \Big\} \times 2$ \\
\bottomrule
\end{tabular}
\caption{Detailed architectural specifications of the Encoder part of a single-stage MAXIM backbone. Depth 1-3 denotes Encoder blocks, while depth 4 corresponds to Backbone blocks. Note that in Bottlenecks, we use \texttt{Conv1x1} in RCAB. \textcolor{red}{*} indicates layers that are not employed in the first stage.}
\label{tab:configurations}
\end{table}

The \textcolor{aogreen}{Decoder} part of MAXIM is symmetric with respect to \cref{tab:configurations}, and has the same configuration. For the \textcolor{myyellow}{CGB} necks, we used $b=d=16$ for the depths 1 and 2, while $b=d=8$ is adopted for depth 3. Basically, we set the block and grid sizes as $16$ for high-resolution stages (\ie feature size $\ge 128$) and $8$ for low-resolution stages (\ie feature size $<128$). Consequently, the input images need to have both dimensions to be divisible by 64, requiring the images to be padded by a multiplier of 64 during the inference.

\subsubsection{Comparison with Other MLPs}
\label{ssec:comparison-mlps}



\begin{figure}[!t]
 \centering
 \includegraphics[width=0.96\linewidth]{ 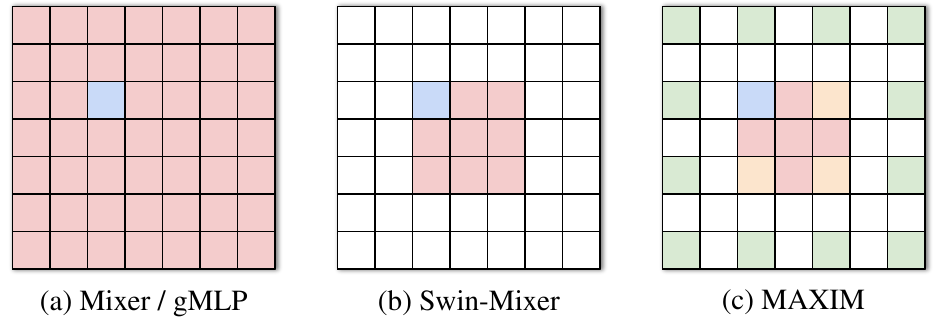}
 \caption{Visualizations of effective receptive fields (shaded area) of the \textcolor{brandeisblue}{blue pixel} for (a) Mixer/gMLP, (b) Swin-Mixer, and (c) our MAXIM. MAXIM attains both local (\textcolor{brickred}{red}) and (dilated) global (\textcolor{aogreen}{green}) perception. \textcolor{myyellow}{Yellow} pixels are achievable by both local and global branches.}
 \label{fig:receptive-field}
\end{figure}

In \cref{fig:receptive-field}, we show a visual comparison of the approximated effective receptive fields among recent MLP models: MLP-Mixer~\cite{tolstikhin2021mlp}, gMLP~\cite{liu2021pay}, Swin-Mixer~\cite{liu2021Swin}, and our proposed MAXIM. Our approach achieves sparse interactions to obtain both local (\textcolor{brickred}{red} in \cref{fig:receptive-field}c) and global dilated (\textcolor{aogreen}{green}) spatial communications. Moreover, as shown in \cref{tab:comparisons-mlps}, unlike previous MLP models, MAXIM obtains both global and fully-convolutional properties with a linear complexity with respect to the number of pixels $N$.

\subsection{JAX Implementations}
\label{ssec:jax-implementations.}

Here we provide a JAX~\cite{jax2018github} implementation of the key component of MAXIM, namely the multi-axis gated MLP block (MAB), in \cref{alg:mab}.

\subsection{Performance \vs Complexity}
\label{sec:performance-complexity}

We demonstrate the performance \vs complexity trade-off in \cref{tab:model-complexity} as compared with other competing methods for all the tasks. As it can be seen, our model obtains state-of-the-art performance at a very moderate complexity. On denoising, for example, MAXIM-3S has only $21\%$ FLOPs and $70\%$ parameters of MIRNet~\cite{zamir2020learning}; on deblurring, our MAXIM-3S model requires only $25\%$ of the number of parameters of the previous best model HINet~\cite{chen2021hinet}, and merely $19\%$ of the number of parameters of the Transformer model IPT~\cite{chen2021pre}. It is also worth noting that unlike IPT, our model requires no large-scale pre-training to obtain leading performance, making it attractive for low-level tasks where datasets are often at limited scale.

\begin{table}[!t]
\centering
\footnotesize
\setlength{\tabcolsep}{4pt}
\renewcommand{\arraystretch}{1.1}
\begin{tabular}{lccc}
\toprule
Model  & Complexity & Fully-conv & Global \\
\midrule
 MLP-Mixer~\cite{tolstikhin2021mlp}  & $\mathcal{O}(N^2)$ & \xmark & \cmark  \\
 gMLP~\cite{liu2021pay} & $\mathcal{O}(N^2)$ & \xmark & \cmark \\
 Swin-Mixer~\cite{liu2021Swin} & $\mathcal{O}(N)$ & \cmark & \xmark \\
 \hline
 MAXIM (ours) & $\mathcal{O}(N)$ & \cmark & \cmark \\
\bottomrule
\end{tabular}
\caption{Comparisons of MAXIM with other MLP models. Our model is both fully-convolutional and global, having a linear complexity with respect to the number of pixels $N$.}
\label{tab:comparisons-mlps}
\end{table}

\subsection{Additional Experiments}
\label{sec:additional-experiments}

Due to limited space in the main paper, we also show experimental results on deblurring and raindrop removal.

\noindent\textbf{Deblurring on REDS}~\cite{Nah_2021_CVPR}. \cref{tab:reds-results} shows quantitative comparisons of MAXIM-3S against the winning solution, HINet~\cite{chen2021hinet}, and a leading model, MPRNet~\cite{zamir2021multi} on the REDS dataset of NTIRE 2021 Image Deblurring Challenge Track 2 JPEG artifacts~\cite{Nah_2021_CVPR}. The metrics are computed and averaged on 300 validation images. Our MAXIM-3S model surpasses HINet by \textbf{0.1} dB of PSNR.

\begin{table}[!t]
\centering
\footnotesize
\setlength{\tabcolsep}{4pt}
\begin{tabular}{l|l|l|c|r|r}
\textbf{Task} & \textbf{Dataset} & \textbf{Model} & \textbf{PSNR} & \textbf{Params} & \textbf{FLOPs} \\ 
\toprule
\multirow{3}{*}{Denoise} & \multirow{3}{*}{SIDD~\cite{abdelhamed2018high}} & MPRNet~\cite{zamir2021multi} & 39.71 & 15.7M & 1176G \\
& &  MIRNet~\cite{zamir2020learning} & 39.72 & 31.7M & 1572G \\
& & \textbf{MAXIM-3S} & 39.96 & 22.2M & 339G \\
\hline
\multirow{4}{*}{Deblur} & \multirow{4}{*}{GoPro~\cite{nah2017deep}} & MPRNet~\cite{zamir2021multi} & 32.66 & 20.1M & 1554G \\
& & HINet~\cite{chen2021hinet} & 32.71 & 88.7M & 341G \\
& & IPT~\cite{chen2021pre} & 32.58 &  114M & 1188G \\
& & \textbf{MAXIM-3S} & 32.86 & 22.2M & 339G \\
\hline
\multirow{3}{*}{Derain} & \multirow{3}{*}{\begin{tabular}{@{}c@{}}Rain13k\\ (Average)\end{tabular}} & MSPFN~\cite{jiang2020multi} & 30.75 & 21.7M & - \\
& & MPRNet~\cite{zamir2021multi} & 32.73 &  3.64M & 297G \\
& & \textbf{MAXIM-2S} & 33.24 & 14.1M & 216G \\
\hline
\multirow{3}{*}{Dehaze} & \multirow{3}{*}{Indoor~\cite{li2019benchmarking}} & MSBDN~\cite{dong2020multi} & 33.79 & 31.3M & 83G \\
& & FFA-Net~\cite{qin2020ffa} & 36.36 & 4.5M & 576G \\
& & \textbf{MAXIM-2S}  & 39.72 & 14.1M & 216G \\
\hline
\multirow{2}{*}{Enhance} & \multirow{2}{*}{LOL~\cite{wei2018deep}} & MIRNet~\cite{zamir2020learning} & 24.14 & 31.7M & 1572G \\
& & \textbf{MAXIM-2S}  & 23.43 & 14.1M & 216G \\

\bottomrule
\end{tabular}
\caption{Model performance \vs complexity comparison of our model with other competing methods for all the tasks. FLOPs are calculated on an input image of size $256\times 256$.}
\label{tab:model-complexity}
\end{table}

\begin{table}[!t]
\footnotesize
\setlength{\tabcolsep}{6pt}
    \centering
    \begin{tabular}{l|cc}
    & \multicolumn{2}{c}{REDS~\cite{Nah_2021_CVPR}} \\
    \toprule
    Method    & PSNR & SSIM  \\
    \midrule
    MPRNet~\cite{zamir2021multi} & 28.79 & 0.911 \\
    HINet~\cite{chen2021hinet} & \underline{28.83} & \underline{0.862} \\
    \midrule
    MAXIM-3S & \textbf{28.93} & \textbf{0.865} \\
    \bottomrule
    \end{tabular}
    \caption{Deblurring comparisons on REDS. Our method outperforms previous winning solution (HINet) on the REDS dataset of NTIRE 2021 Image Deblurring Challenge Track 2 JPEG artifacts. The scores are evaluated on 300 images from the validation set. Results are gathered from the authors of~\cite{chen2021hinet}.}
    \label{tab:reds-results}
\end{table}

\begin{table}[!t]
\footnotesize
\setlength{\tabcolsep}{4pt}
    \centering
    \begin{tabular}{l|cc|cc}
    & \multicolumn{2}{c}{Raindrop-A~\cite{qian2018attentive}} & \multicolumn{2}{c}{Raindrop-B~\cite{qian2018attentive}} \\
    \toprule
    Method    & PSNR & SSIM  & PSNR & SSIM  \\
    \midrule
    AGAN~\cite{qian2018attentive} & \underline{31.62} & 0.921 & 25.05 & 0.811 \\
    DuRN~\cite{liu2019dual} & 31.24 & 0.926 & \underline{25.32} & \underline{0.817} \\
    Quan~\cite{quan2019deep} & 31.36 & \underline{0.928} & - & - \\
    \midrule
    MAXIM-2S & \textbf{31.87} & \textbf{0.935} & \textbf{25.74} & \textbf{0.827} \\
    \bottomrule
    \end{tabular}
    \caption{Deraining comparisons on Raindrop removal dataset~\cite{qian2018attentive}. Our MAXIM-2S model attains state-of-the-art performance on both Raindrop testset A and B.}
    \label{tab:raindrop-results}
\end{table}

\noindent\textbf{Raindrop removal}~\cite{qian2018attentive}. Apart from the rain streak removal task reported in the main paper, we also evaluated our MAXIM model on the raindrop removal task. As can be seen in \cref{tab:raindrop-results}, our model achieved the best performance: \textbf{31.87} dB and \textbf{25.74} dB PSNR on Raindrop testset A and B.

\subsection{More Visual Comparisons}
\label{sec:more-visual-comparisons}

\begin{figure*}[!htb]
    \center \includegraphics[width=0.96\linewidth]{ 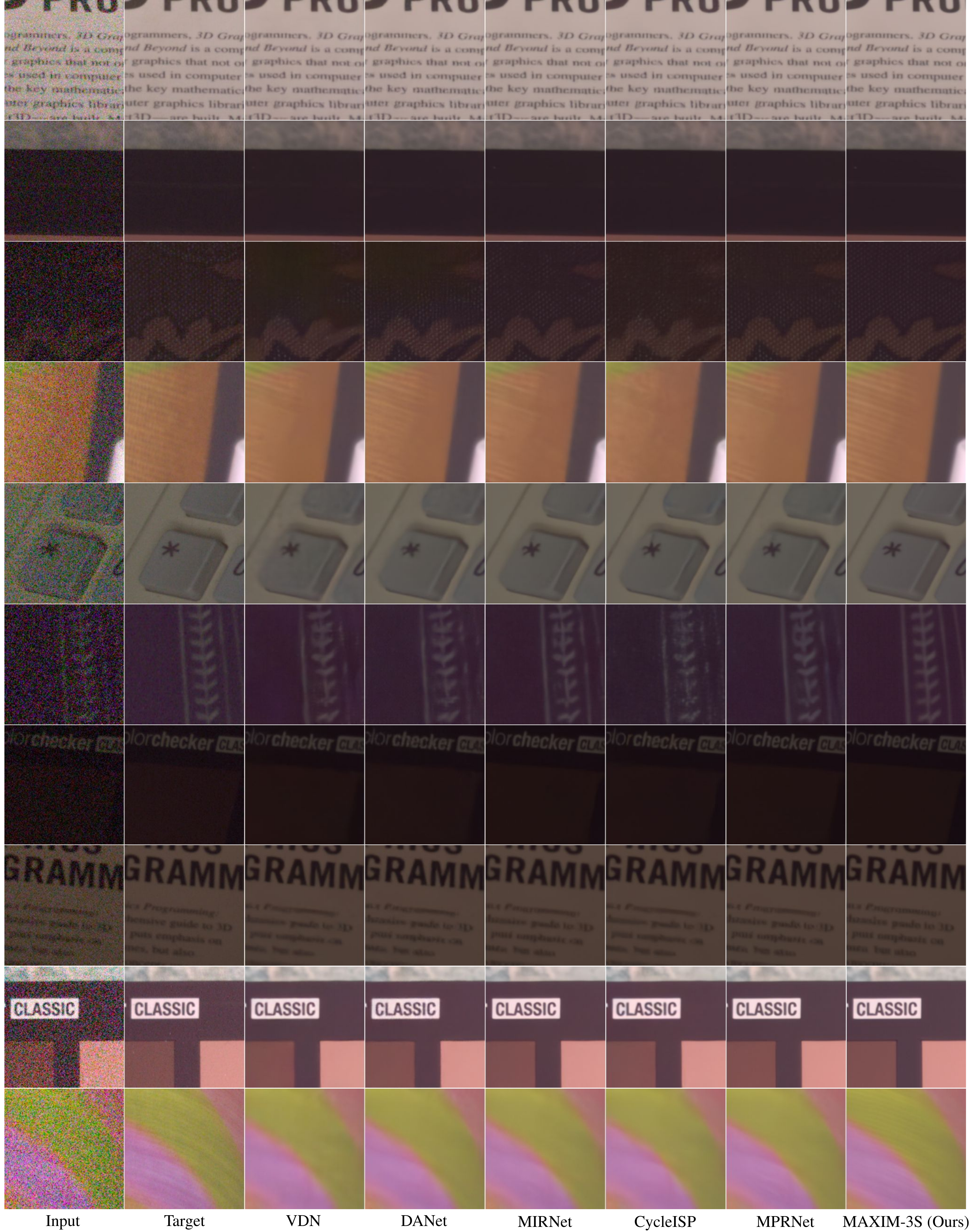}
\caption{Visual examples for image denoising on SIDD~\cite{abdelhamed2018high} among VDN~\cite{yue2019variational}, DANet~\cite{yue2020dual}, MIRNet~\cite{zamir2020learning}, CycleISP~\cite{zamir2020cycleisp}, MPRNet~\cite{zamir2021multi}, and the proposed MAXIM-3S. Our model clearly removed real noise while recovering more details.}
\label{fig:sidd-visual}
\end{figure*}

\noindent\textbf{Denoising.} \cref{fig:sidd-visual} shows denoising results of our model compared with SOTA models on SIDD~\cite{abdelhamed2018high}. Our model recovers more details, yielding visually pleasant outputs.

\begin{figure*}[!htb]
    \center \includegraphics[width=0.90\linewidth]{ 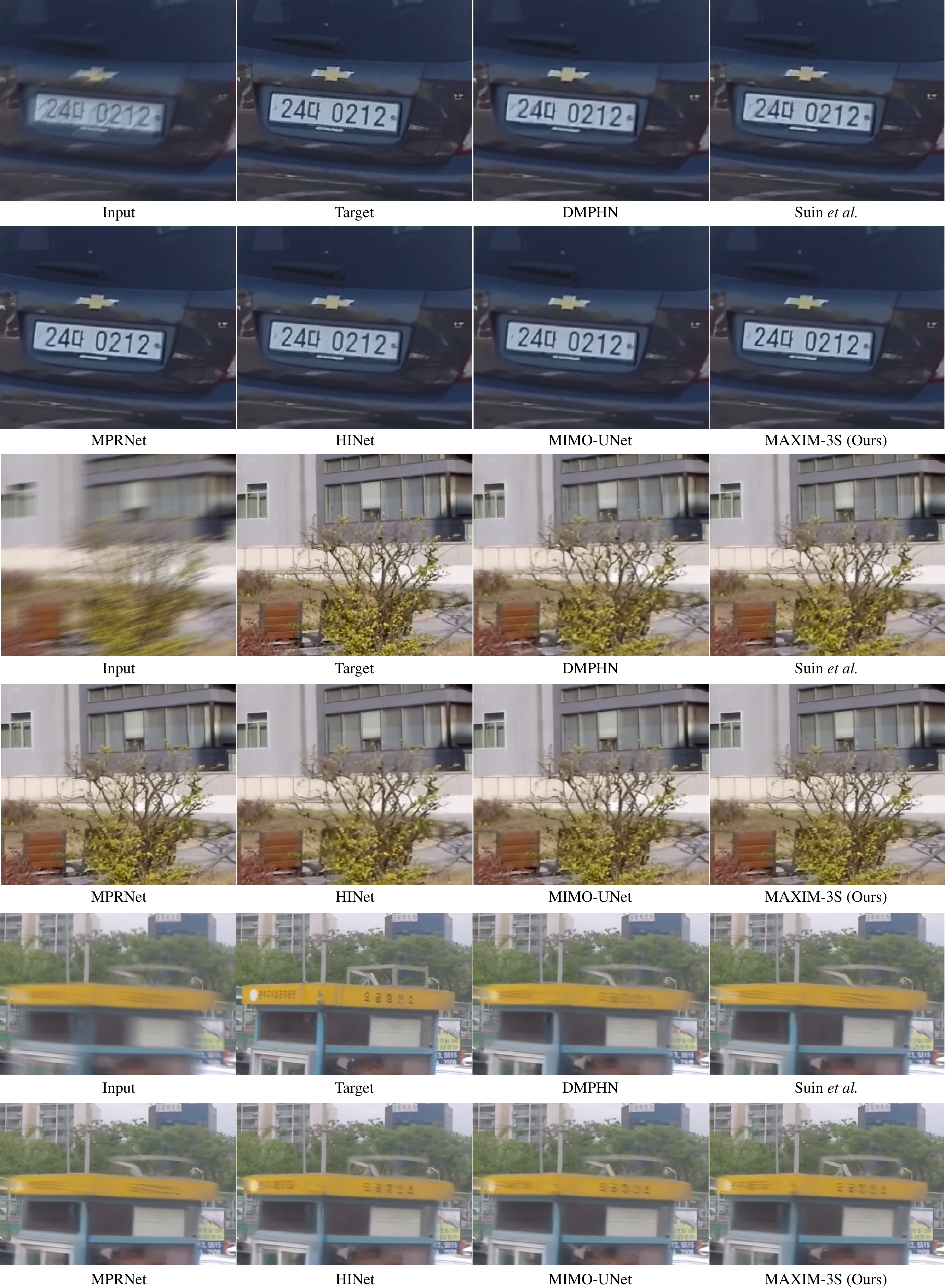}
\caption{Visual examples for image deblurring on GoPro~\cite{nah2017deep} among DMPHN~\cite{zhang2019deep}, Suin \etal~\cite{suin2020spatially}, MPRNet~\cite{zamir2021multi}, HINet~\cite{chen2021hinet}, MIMO-UNet~\cite{cho2021rethinking}, and our MAXIM-3S.}
\label{fig:gopro-visual}
\end{figure*}

\begin{figure*}[!htb]
    \center \includegraphics[width=0.95\linewidth]{ 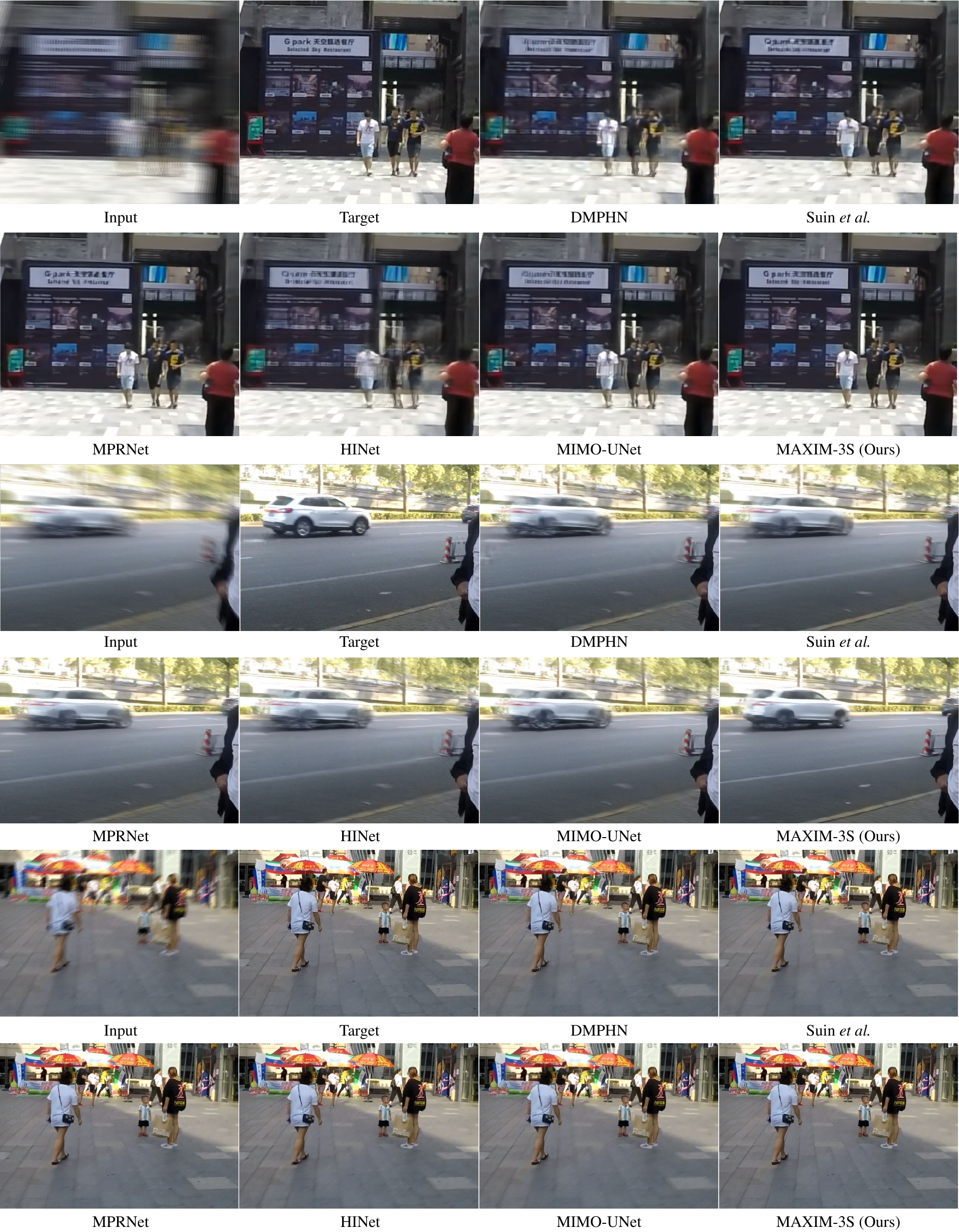}
\caption{Visual comparisons for image deblurring on HIDE~\cite{shen2019human} among DMPHN~\cite{zhang2019deep}, Suin \etal~\cite{suin2020spatially}, MPRNet~\cite{zamir2021multi}, HINet~\cite{chen2021hinet}, MIMO-UNet~\cite{cho2021rethinking}, and our MAXIM-3S.}
\label{fig:hide-visual}
\end{figure*}

\begin{figure*}[!htb]
    \center \includegraphics[width=0.99\linewidth]{ 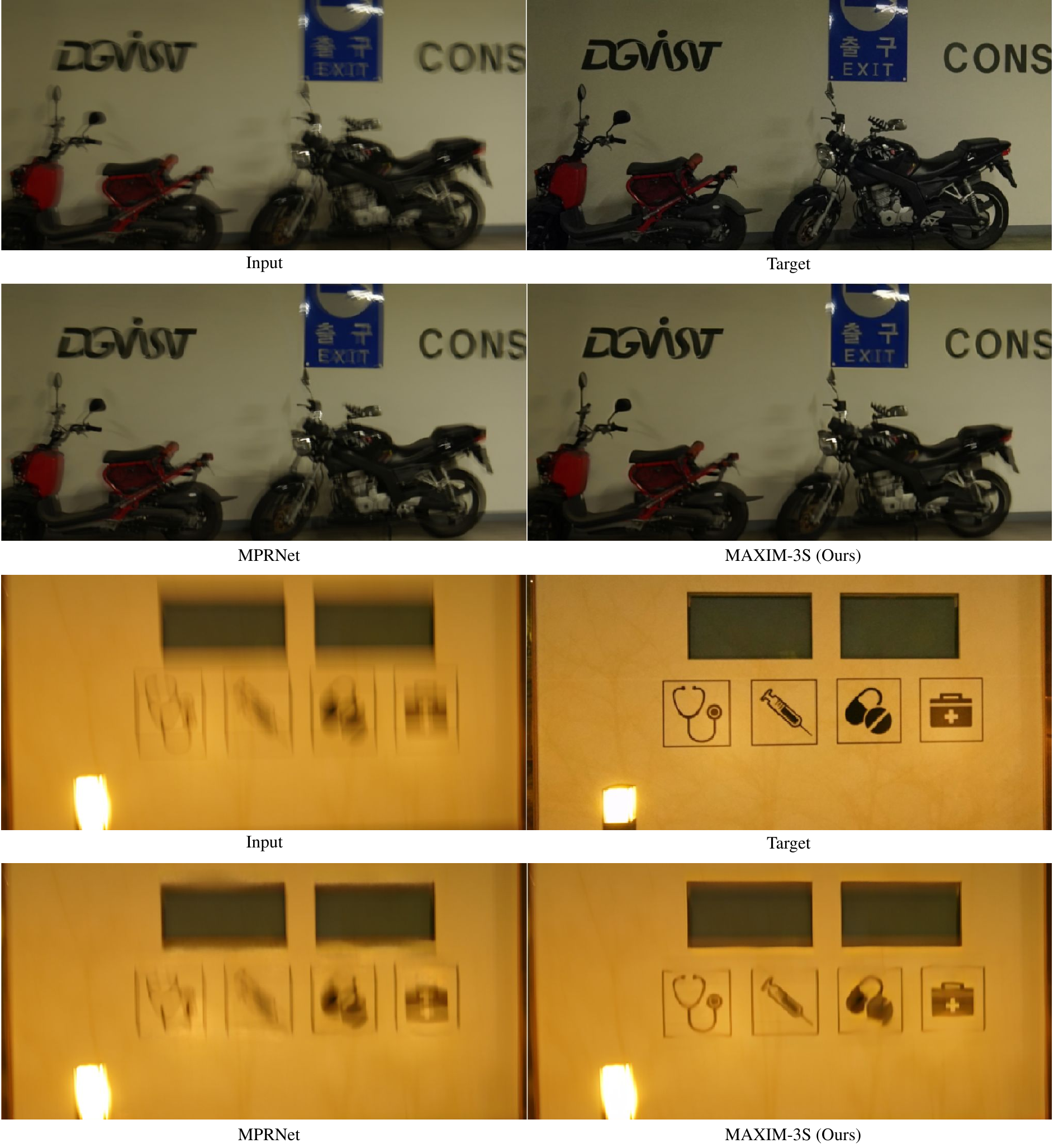}
\caption{Visual comparisons for image deblurring on RealBlur-J~\cite{rim2020real} between previous best model MPRNet~\cite{zamir2021multi} and MAXIM-3S.}
\label{fig:realblurj-visual}
\end{figure*}

\begin{figure*}[!htb]
    \center \includegraphics[width=0.99\linewidth]{ 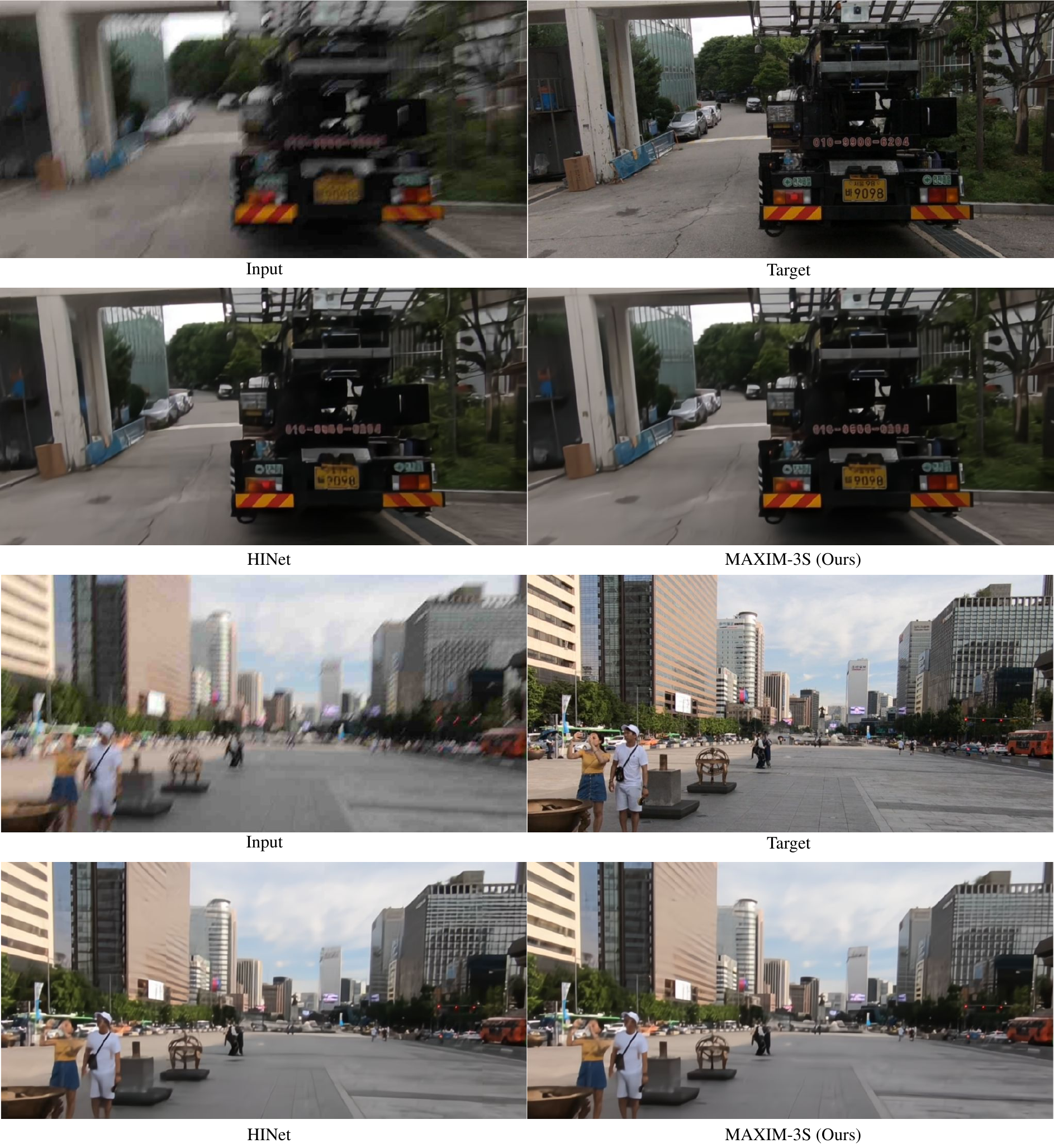}
\caption{Visual comparisons for image deblurring on REDS~\cite{Nah_2021_CVPR} between our model and the winning solution, HINet~\cite{chen2021hinet}, for REDS dataset of the NTIRE 2021 Image Delurring Challenge Track 2 JPEG artifacts~\cite{Nah_2021_CVPR}.}
\label{fig:reds-visual}
\end{figure*}

\noindent\textbf{Deblurring.} The visual results on GoPro~\cite{nah2017deep}, HIDE~\cite{shen2019human}, RealBlur-J~\cite{rim2020real}, and REDS~\cite{Nah_2021_CVPR} are shown in \cref{fig:gopro-visual}, \cref{fig:hide-visual}, \cref{fig:realblurj-visual}, and \cref{fig:reds-visual}, respectively. Our model outperformed other competing methods on both synthetic and real-world deblurring benchmarks.

\begin{figure*}[!htb]
    \center \includegraphics[width=0.92\linewidth]{ 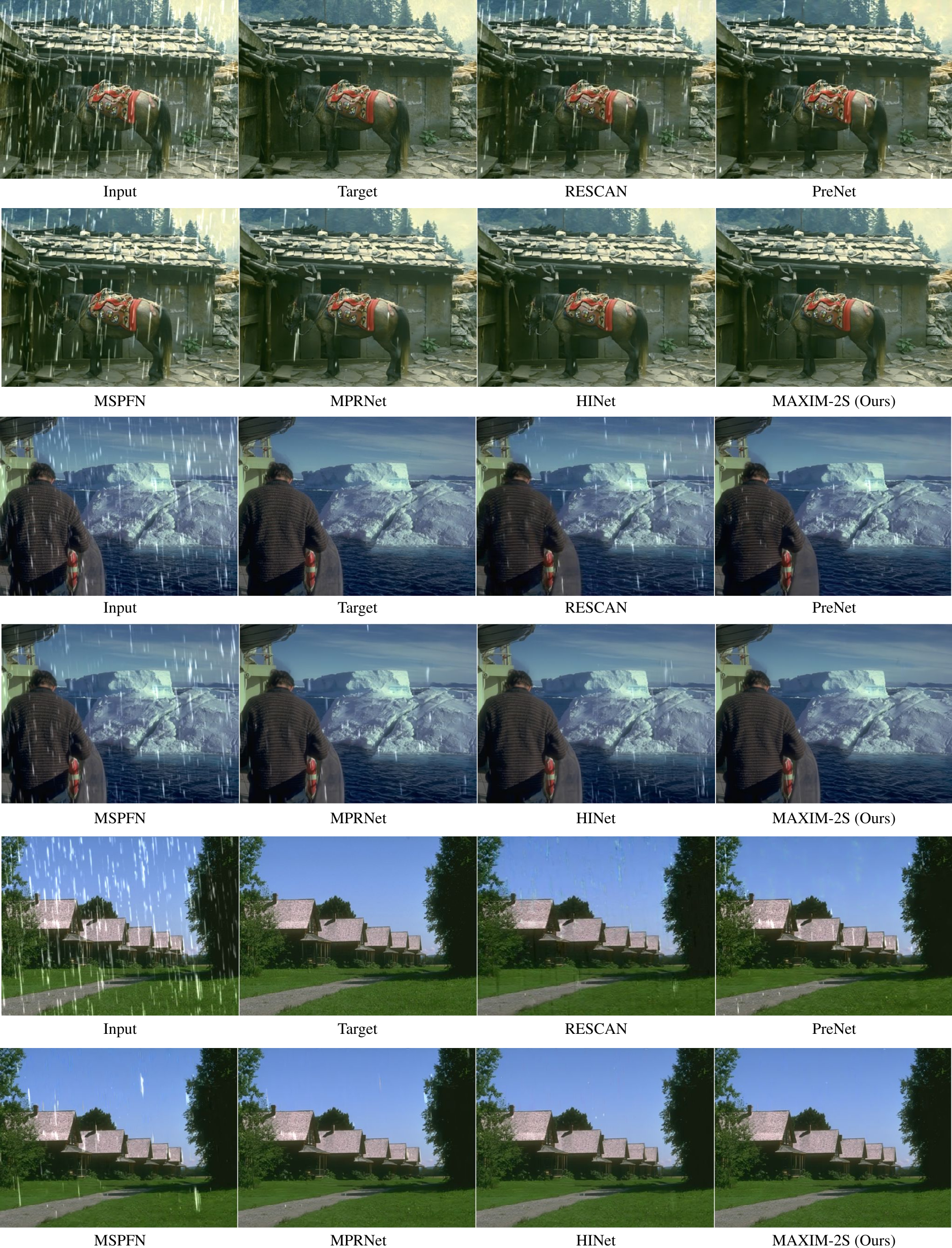}
\caption{Visual examples for image deraining on Rain100L~\cite{yang2017deep} among RESCAN~\cite{li2018recurrent}, PreNet~\cite{ren2019progressive}, MSPFN~\cite{jiang2020multi}, MPRNet~\cite{zamir2021multi}, HINet~\cite{chen2021hinet}, and our MAXIM-2S model.}
\label{fig:rain100l-visual}
\end{figure*}

\begin{figure*}[!htb]
    \center \includegraphics[width=0.96\linewidth]{ 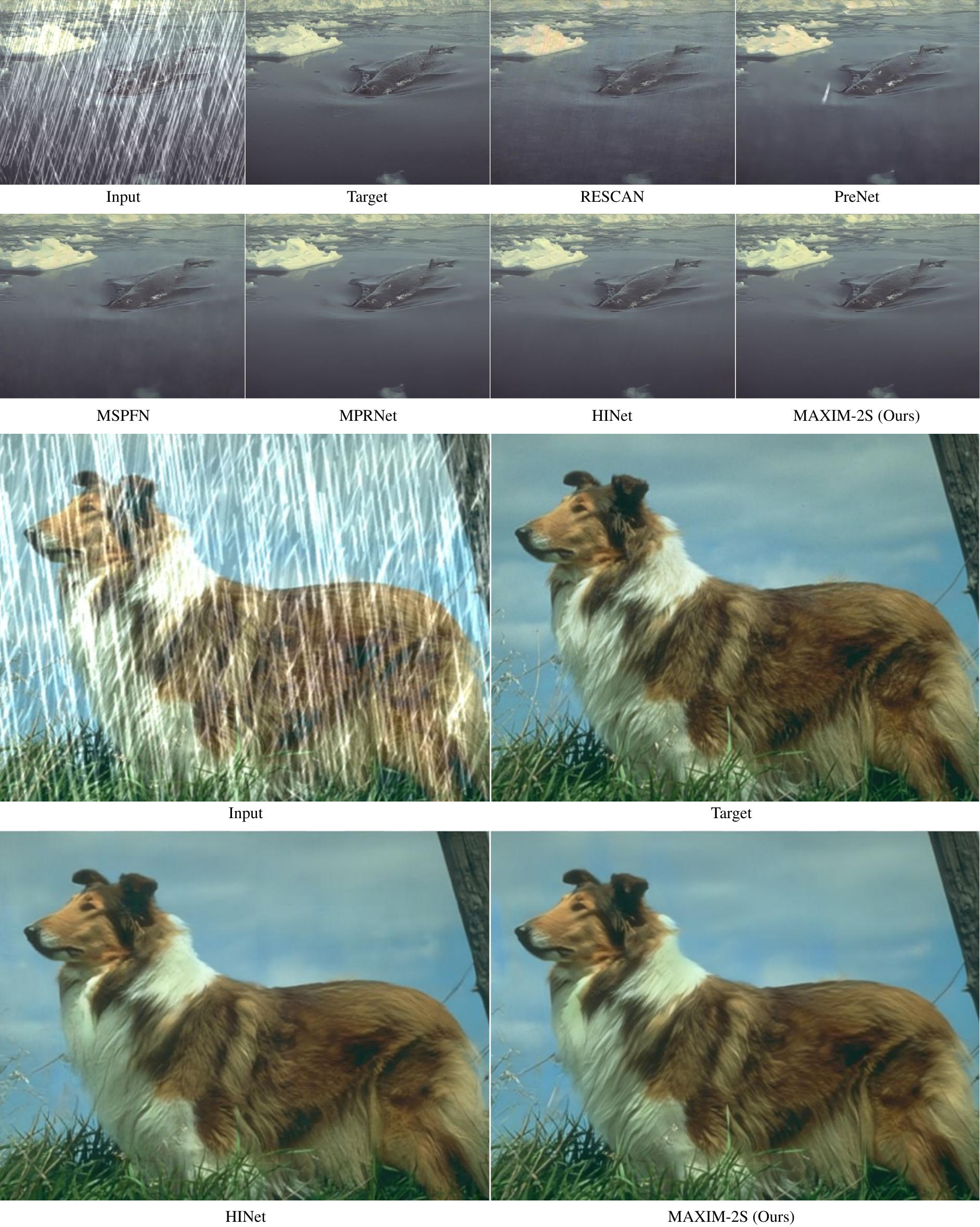}
\caption{Visual examples for image deraining on Rain100H~\cite{yang2017deep}. At extremely high raining levels, our model recovers more details and textures compared to previous competitive methods.}
\label{fig:rain100h-visual}
\end{figure*}

\begin{figure*}[!htb]
    \center \includegraphics[width=0.99\linewidth]{ 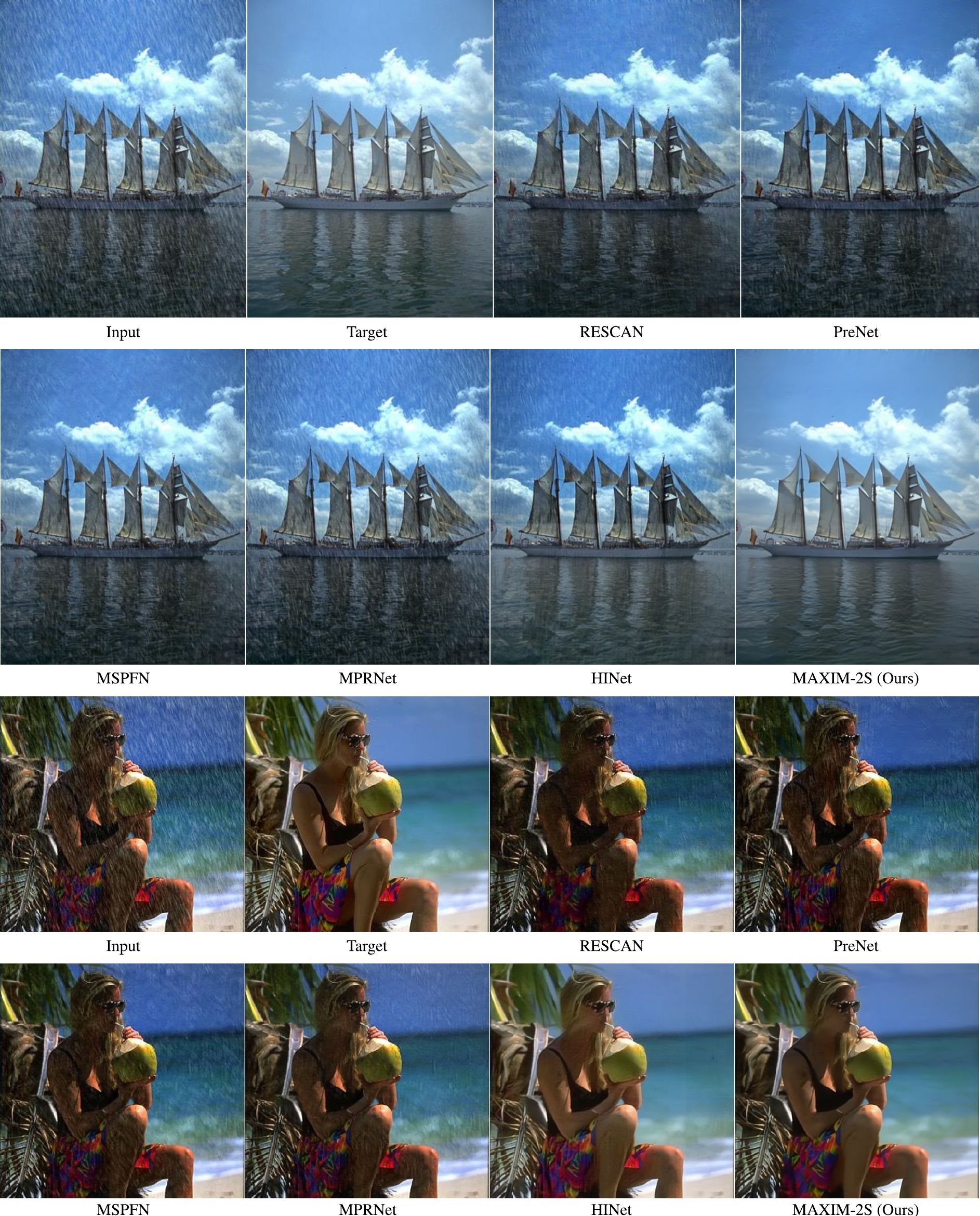}
\caption{Visual examples for image deraining on Test100~\cite{zhang2019image}. Our model removes both raining streaks and visible JPEG artifacts.}
\label{fig:test100-visual}
\end{figure*}

\begin{figure*}[!htb]
    \center \includegraphics[width=0.99\linewidth]{ 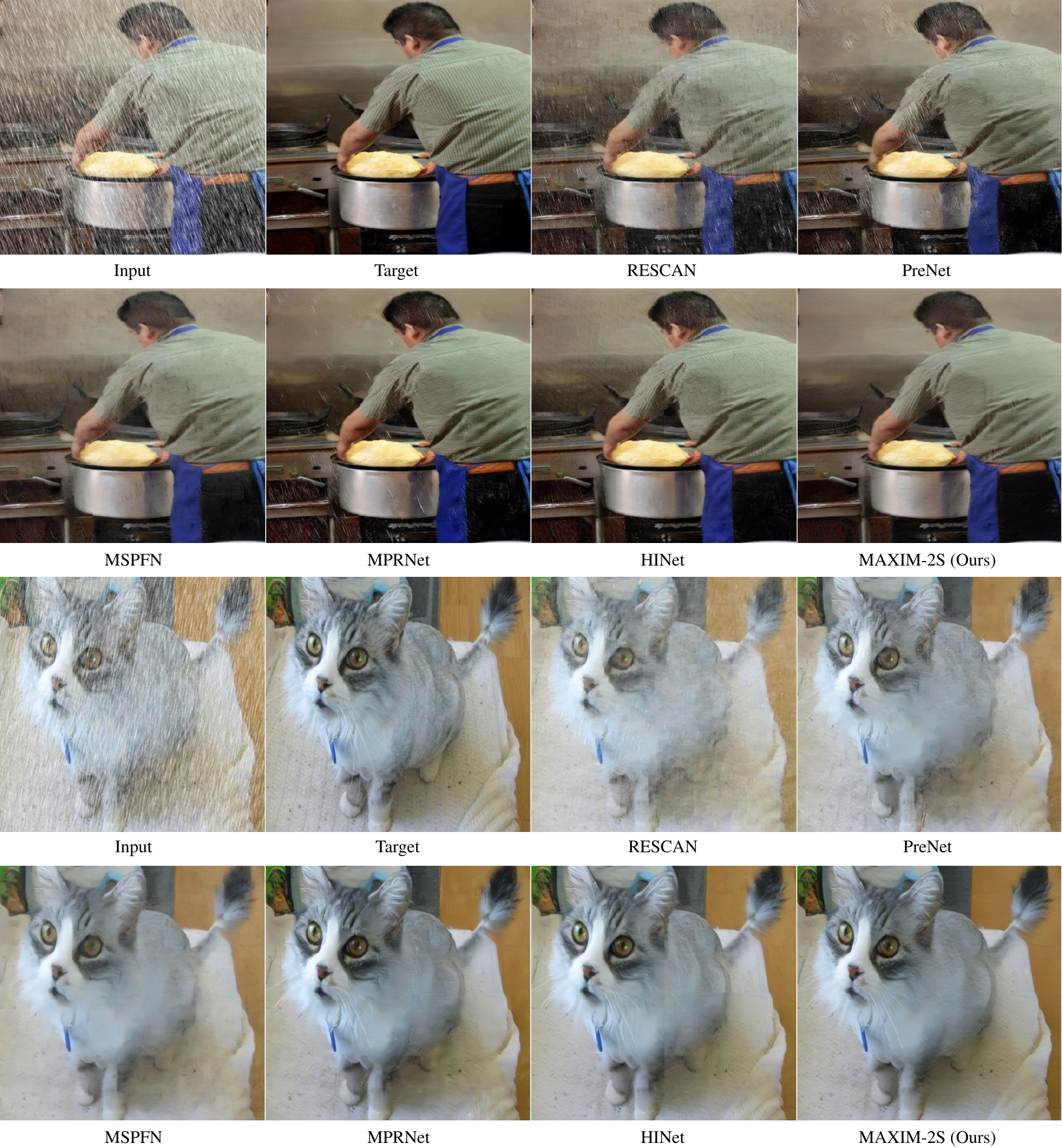}
\caption{Visual examples for image deraining on Test1200~\cite{zhang2018density}.}
\label{fig:test1200-visual}
\end{figure*}

\noindent\textbf{Deraining.} Qualitative comparisons of our model against SOTA methods on deraining are shown in \cref{fig:rain100l-visual}, \cref{fig:rain100h-visual}, \cref{fig:test100-visual}, and \cref{fig:test1200-visual}.

\begin{figure*}[!htb]
    \center \includegraphics[width=0.99\linewidth]{ 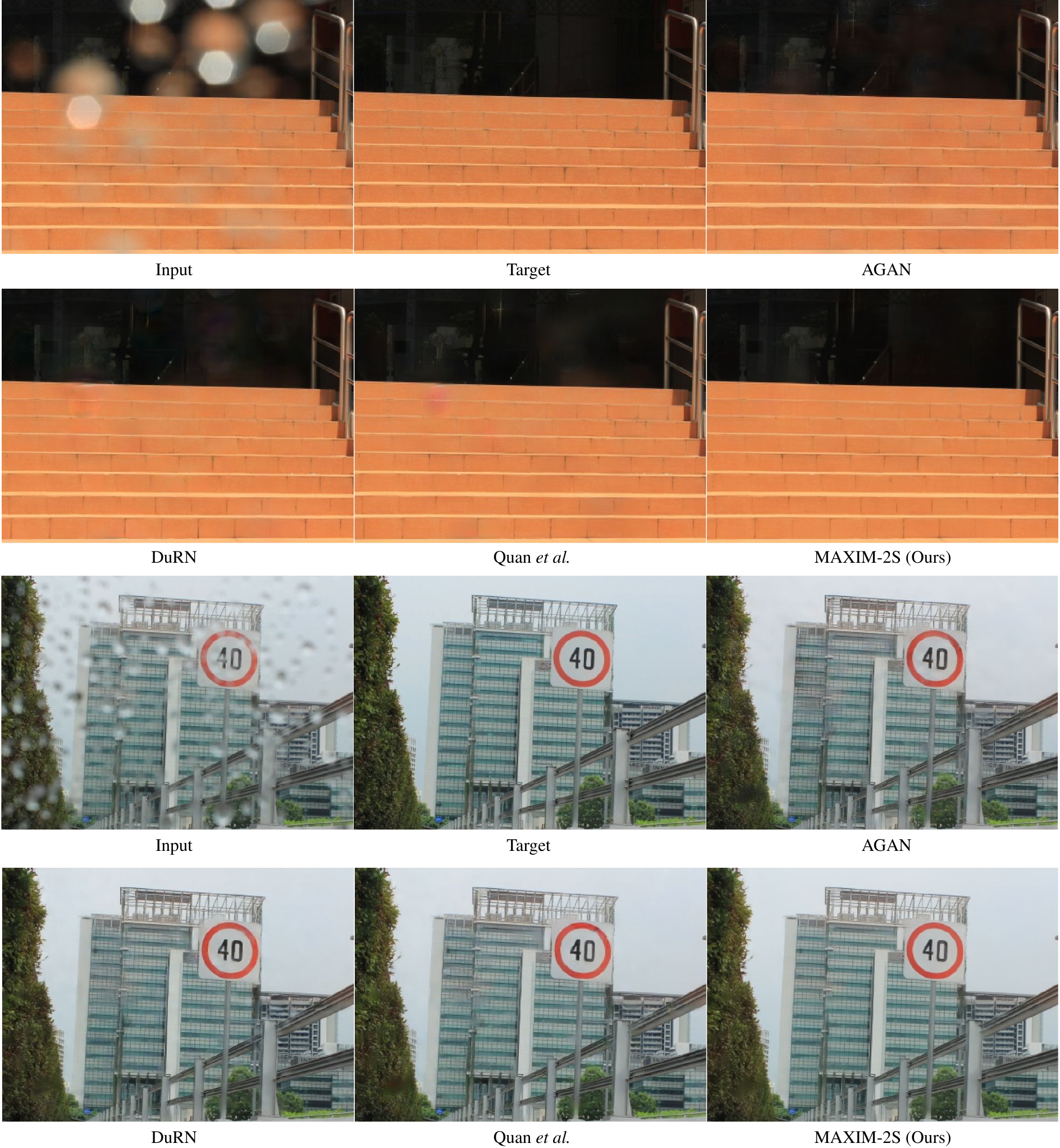}
\caption{Visual comparisons for raindrop removal on Raindrop-A~\cite{qian2018attentive} among  AGAN~\cite{qian2018attentive}, DuRN~\cite{liu2019dual}, Quan~\cite{quan2019deep}, and MAXIM-2S.}
\label{fig:raindropa-visual}
\end{figure*}

\begin{figure*}[!htb]
    \center \includegraphics[width=0.99\linewidth]{ 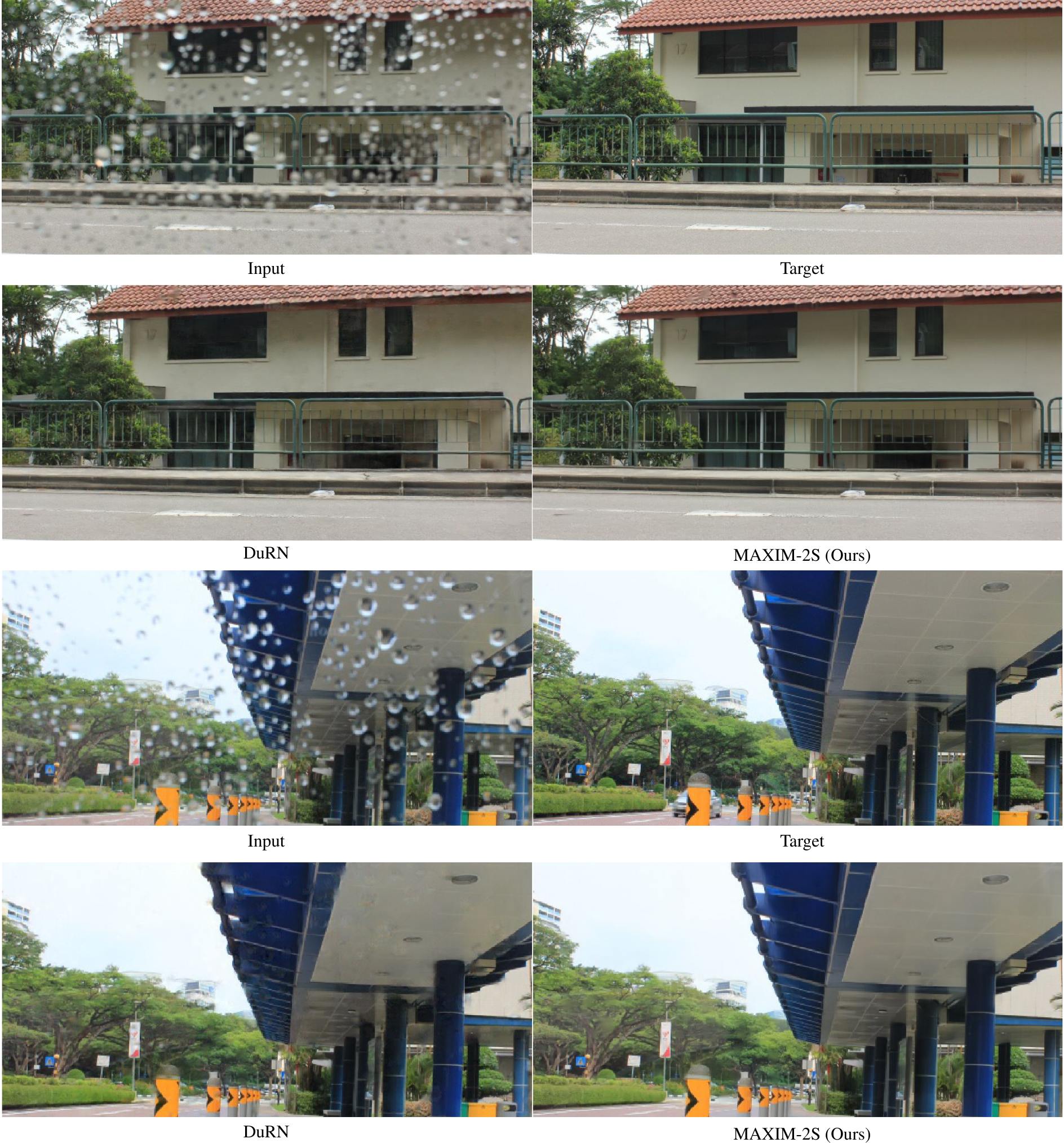}
\caption{Visual comparisons for raindrop removal on Raindrop testset B~\cite{qian2018attentive}.}
\label{fig:raindropb-visual}
\end{figure*}

\noindent\textbf{Raindrop removal.} We provide visual comparisons of the raindrop removal task on the Raindrop testset A and B~\cite{qian2018attentive} in \cref{fig:raindropa-visual} and \cref{fig:raindropb-visual}. 

\begin{figure*}[!htb]
    \center \includegraphics[width=0.82\linewidth]{ 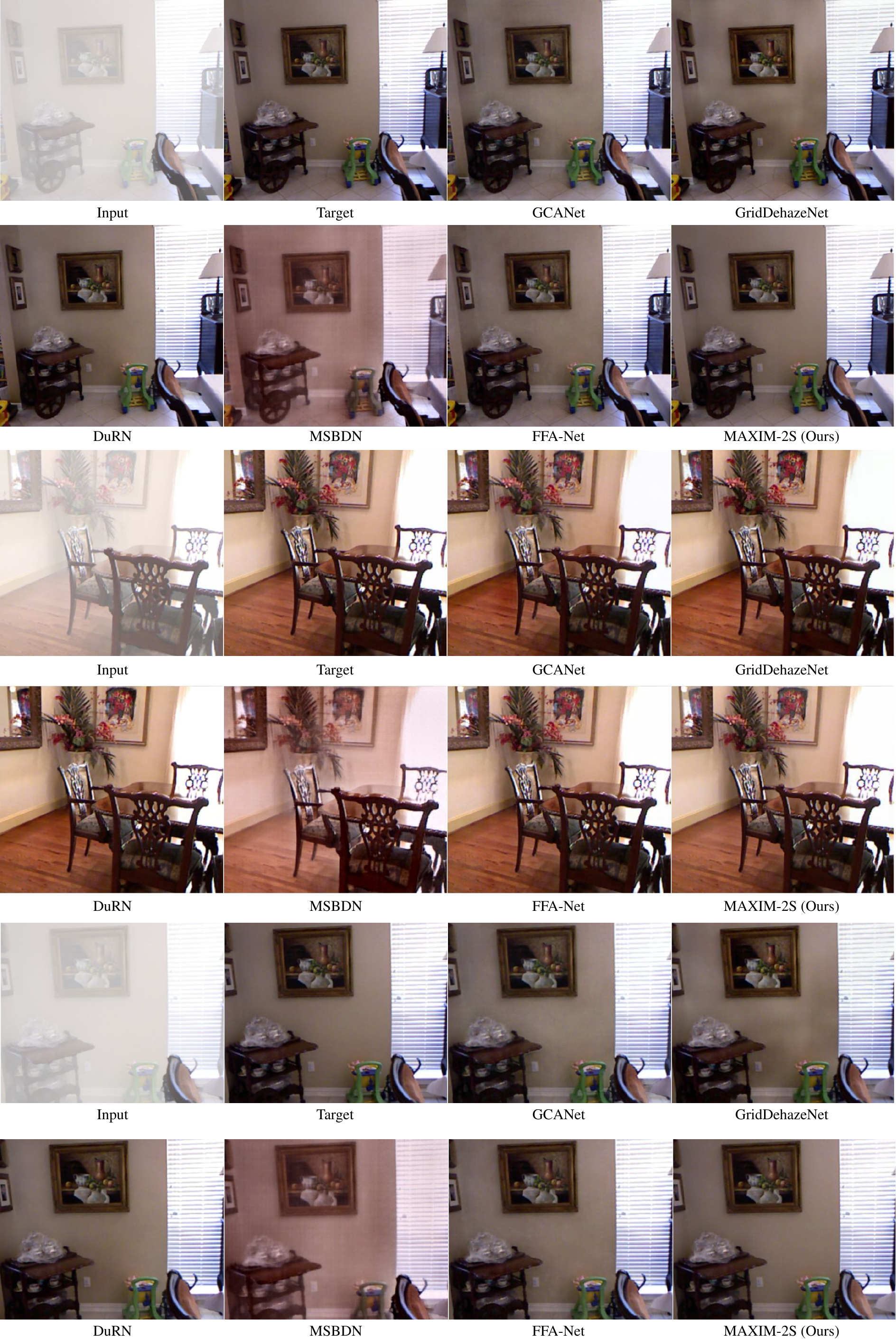}
\caption{Visual comparisons for image dehazing on SOTS indoor testset~\cite{li2019benchmarking} among GCANet~\cite{chen2019gated}, GridDehaze~\cite{liu2019griddehazenet}, DuRN~\cite{liu2019dual}, MSBDN~\cite{dong2020multi}, FFA-Net~\cite{qin2020ffa}, and our MAXIM-2S.}
\label{fig:reside-indoor}
\end{figure*}

\begin{figure*}[!htb]
    \center \includegraphics[width=0.76\linewidth]{ 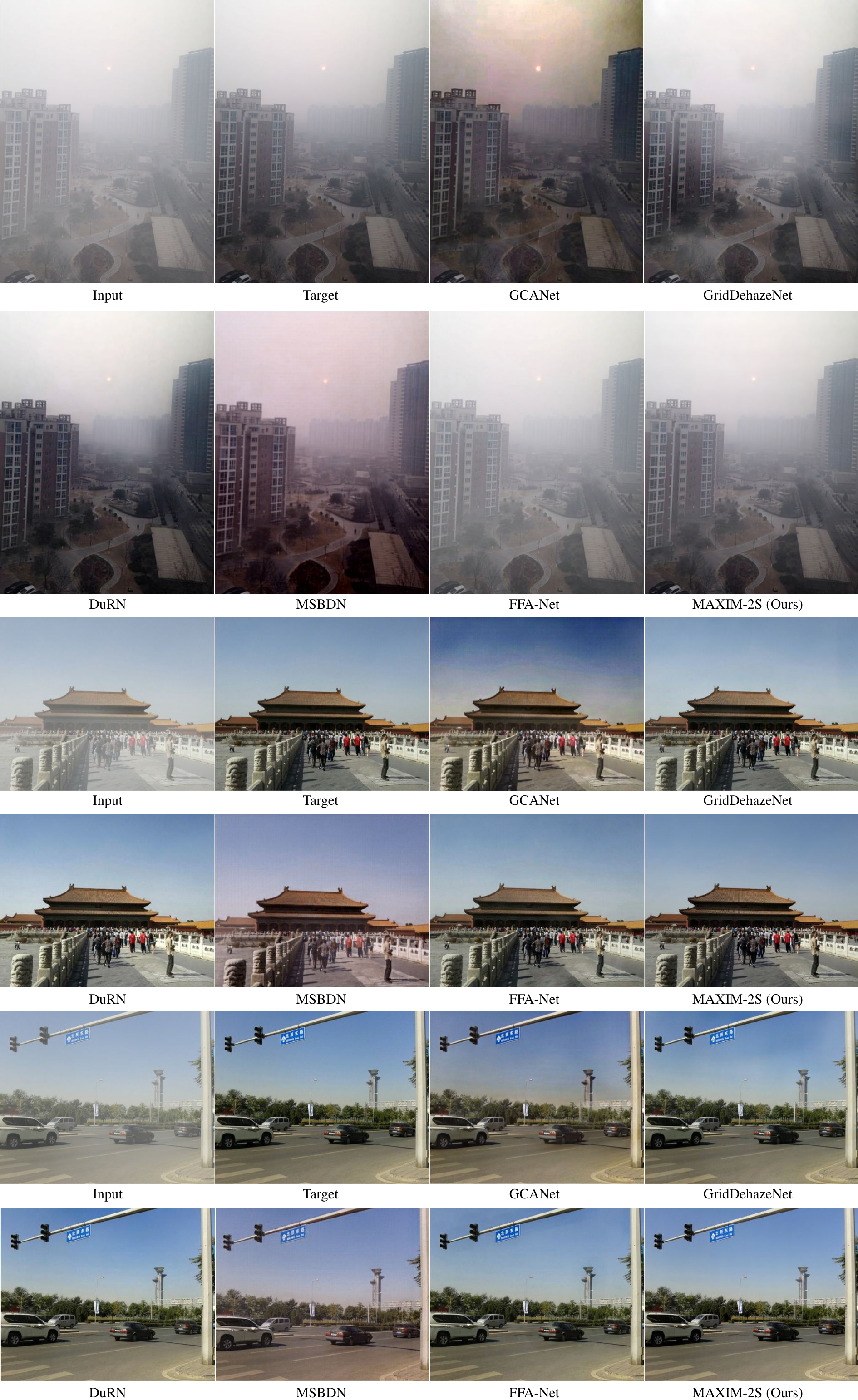}
\caption{Visual comparisons for image dehazing on SOTS outdoor testset~\cite{li2019benchmarking} of MAXIM-2S against other approaches.}
\label{fig:reside-outdoor}
\end{figure*}

\noindent\textbf{Dehazing.} We provide dehazing comparisons on the SOTS~\cite{li2019benchmarking} indoor and outdoor sets in \cref{fig:reside-indoor} and \cref{fig:reside-outdoor}.

\begin{figure*}[!htb]
    \center \includegraphics[width=0.8\linewidth]{ 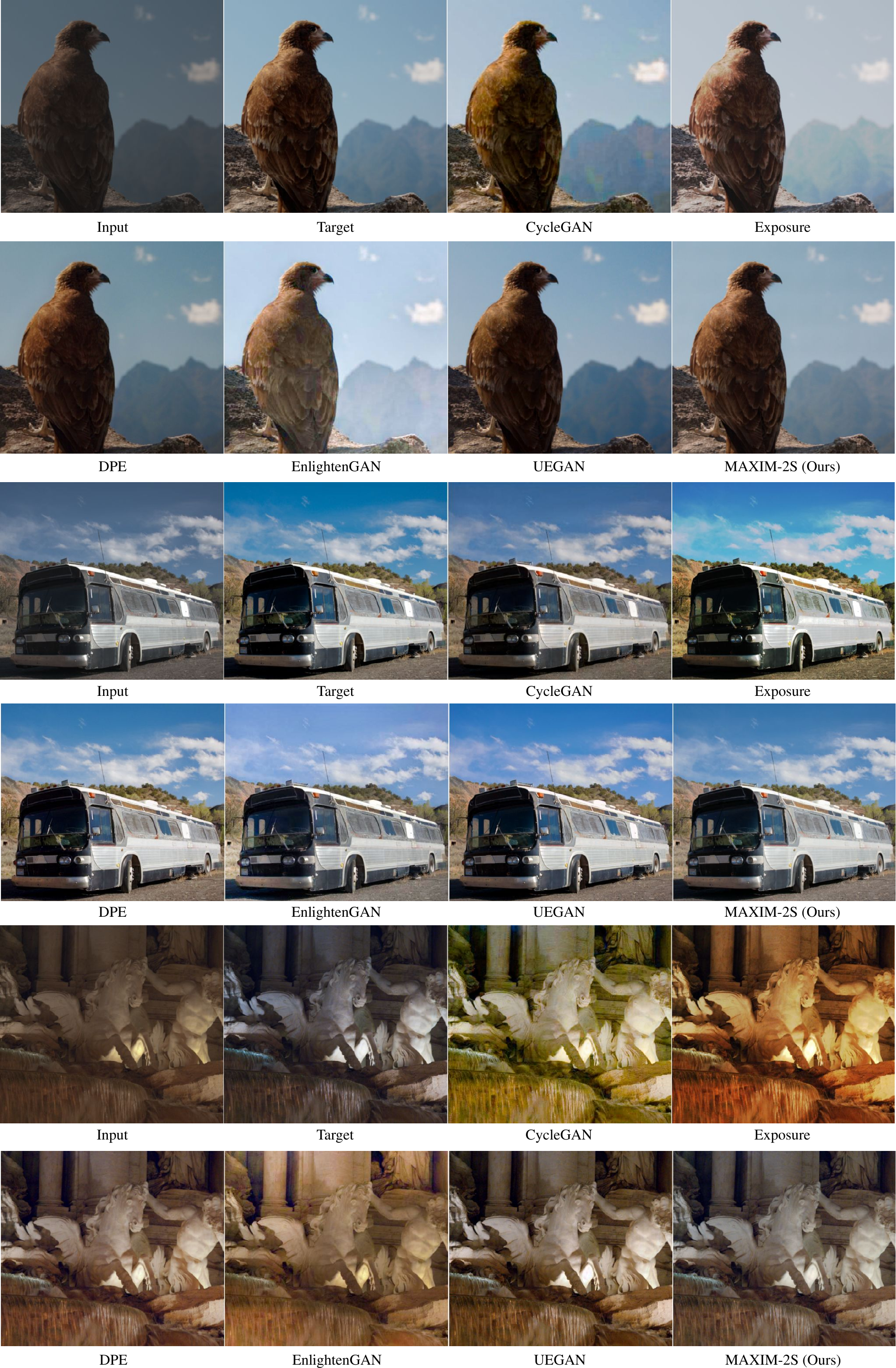}
\caption{Visual comparisons for image retouching on MIT-Adobe FiveK~\cite{bychkovsky2011learning} provided by the authors of~\cite{ni2020towards} among CycleGAN~\cite{zhu2017unpaired}, Exposure~\cite{hu2018exposure}, DPE~\cite{chen2018deep}, EnlightenGAN~\cite{jiang2021enlightengan}, UEGAN~\cite{ni2020towards} and MAXIM-2S.}
\label{fig:fivek-visual}
\end{figure*}

\noindent\textbf{Retouching.} \cref{fig:fivek-visual} shows additional comparisons of our model with competing methods on the Five-K dataset~\cite{bychkovsky2011learning} provided by~\cite{ni2020towards} for retouching results.

\begin{figure*}[!htb]
    \center \includegraphics[width=0.88\linewidth]{ 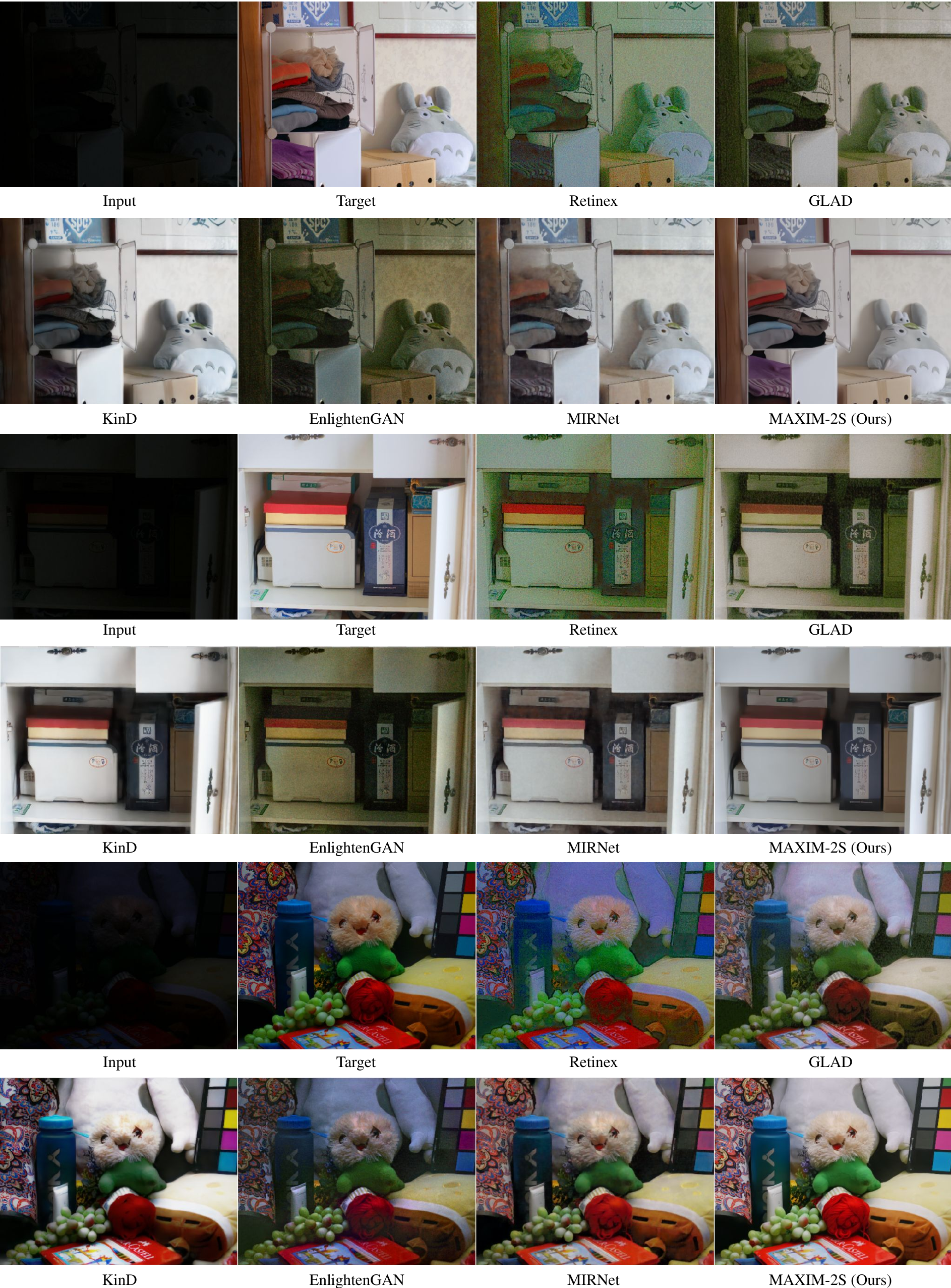}
\caption{Visual examples for image low-light enhancement on the LOL dataset~\cite{wei2018deep} between Retinex~\cite{wei2018deep}, GLAD~\cite{wang2018gladnet}, KinD~\cite{zhang2019kindling}, EnlightenGAN~\cite{jiang2021enlightengan},  MIRNet~\cite{zamir2020learning}, and MAXIM-2S. Our model effectively enhances lighting while largely reducing noise, producing higher-quality images compared to other approaches.}
\label{fig:lol-visual}
\end{figure*}

\noindent\textbf{Low-light enhancement.} \cref{fig:lol-visual} demonstrates the evaluations on the LOL~\cite{wei2018deep} test set for low-light enhancement.

\subsection{Weight Visualizations}
\label{sec:visualization}

\begin{figure*}[!ht]
\centering
\includegraphics[width=0.96\linewidth]{ 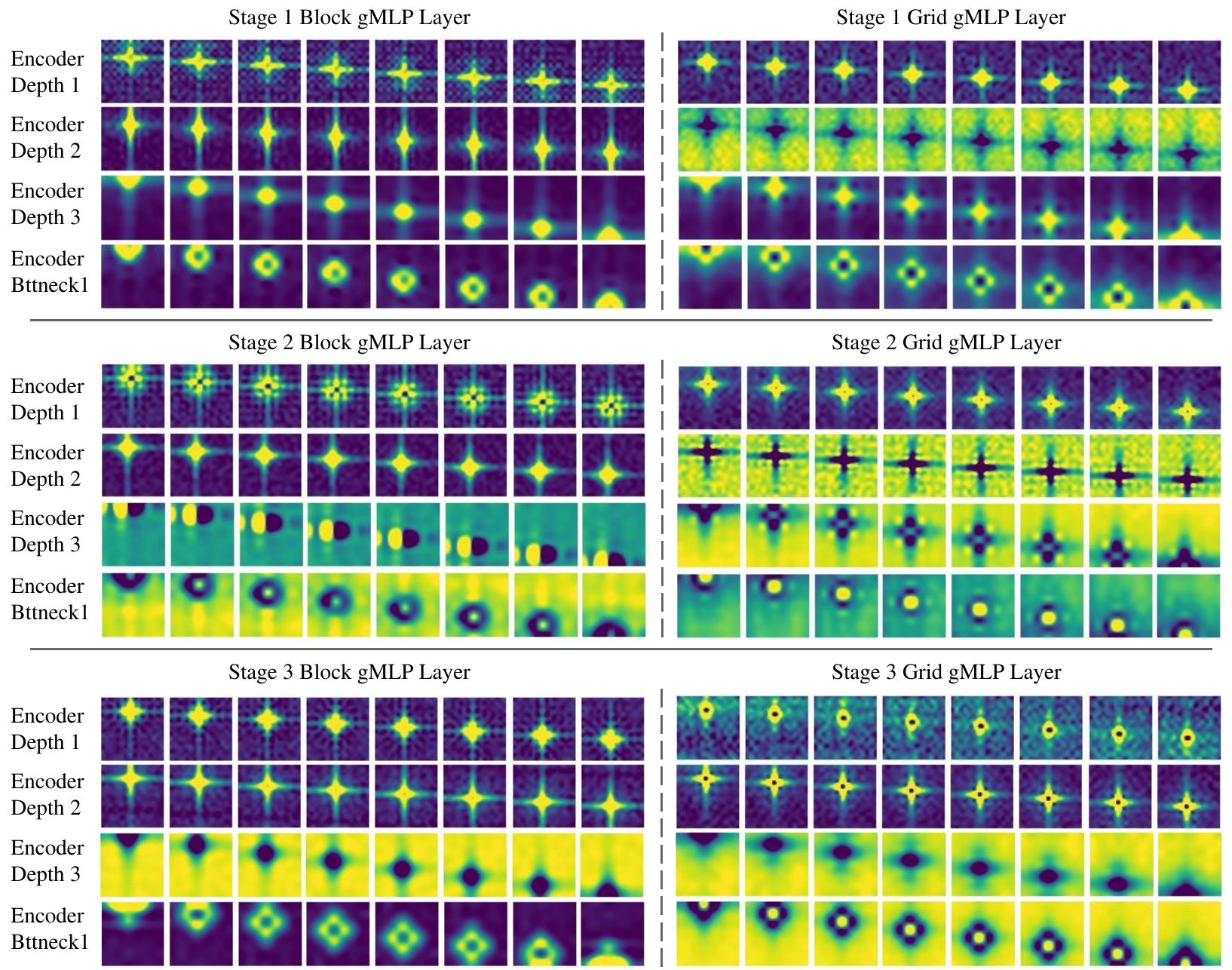}
\caption{Spatial projection weights in block gMLP and grid gMLP layers of the MAXIM-3S model trained on GoPro~\cite{nah2017deep}. Each row shows the filters (reshaped into 2D) for a reduced set of consecutive channels. The filter sizes for Encoder depth 1 and 2 are $16\times 16$, while for Encoder depth 3 and Bottleneck1 are $8\times 8$ (resized to the same shape for better visualization). It is worth noting that the weights of block gMLP layers (left) are directly applied on pixels within local windows and shared at each non-overlapping window of the feature maps (similar to strided convolution), while the weights of grid gMLP layers (right) correspond to a global, dilated aggregation overlaid on the entire image.}
\label{fig:weight-visual}
\end{figure*}

\cref{fig:weight-visual} visualizes the spatial projection matrices of the block gMLP and the grid gMLP layers of each stage of MAXIM-3S trained on GoPro~\cite{nah2017deep}. Similar to~\cite{liu2021pay}, we also observed that the weights after learning exhibit locality and spatial invariance. Surprisingly, the global grid gMLP layer also learns to perform `local' operations (but on the uniform dilated grid). The spatial weights of block gMLP and grid gMLP in the same layer often demonstrate similar or coupled shapes, which may be attributed to the parallel-branch design in the multi-axis gMLP block. However, we have not observed a clear trend on how these filters at different stages vary.

\subsection{Limitations and Discussions}
\label{sec:limitations}

One potential limitation of our model, which is shared with the existing SOTA, is the relatively inadequate generalization to real-world examples. This perhaps can be attributed to the training examples provided by the existing synthesized image restoration benchmarks. Creating more realistic, large-scale datasets through data-generation schemes~\cite{son2021toward,wang2021real} can improve this shortcoming. Also, we observe that our model tends to slightly overfit certain benchmarks, because we did not apply a strong regularization (\eg, dropout) during training. Even though we find that regularization may result in a small reduction in performance for our models on these benchmarks we evaluated, it is worth exploring in future to effectively improve the generalization of our restoration models.

It is worth mentioning that our model is able to generate high quality sharp images, which are visually comparable to the state-of-the-art generative models~\cite{zhao2021improved, jiang2021transgan}. Notably, our model produces more conservative results without hallucinating many nonexistent details, delivering more reliable results than generative models.

\begin{algorithm*}[htbp]
\caption{\small JAX code implementing the Multi-Axis Gated MLP Block (MAB).}
\label{alg:mab}
\definecolor{codeblue}{rgb}{0.25,0.5,0.5}
\definecolor{codekw}{rgb}{0.85, 0.18, 0.50}
\lstset{
  backgroundcolor=\color{white},
  basicstyle=\fontsize{8.5pt}{8.5pt}\ttfamily\selectfont,
  columns=fullflexible,
  breaklines=true,
  captionpos=b,
  stringstyle=\ttfamily\color{red!50!brown},
  commentstyle=\color{codeblue},
  keywordstyle=\color{codekw},
}
\begin{lstlisting}[language=python,escapeinside=``]
from typing import Sequence
import einops
import flax.linen as nn
import jax.numpy as jnp

def block_images(x, patch_size):
  n, h, w, channels = x.shape
  grid_height, grid_width = h // patch_size[0], w // patch_size[1]
  x = einops.rearrange(x, "n (gh fh) (gw fw) c -> n (gh gw) (fh fw) c",
      gh=grid_height, gw=grid_width, fh=patch_size[0], fw=patch_size[1])
  return x

def unblock_images(x, grid_size, patch_size):
  x = einops.rearrange(x, "n (gh gw) (fh fw) c -> n (gh fh) (gw fw) c",
      gh=grid_size[0], gw=grid_size[1], fh=patch_size[0], fw=patch_size[1])
  return x

class SpatialGatingUnit(nn.Module):
  """Gated MLP applied on a specified axis: -3 for grid and -2 for block."""
  @nn.compact
  def __call__(self, x, axis=-3):
    u, v = jnp.split(x, 2, axis=-1)
    v = nn.LayerNorm()(v)
    n = x.shape[axis]   # get spatial dim at the 'grid' or 'block' axis
    v = jnp.swapaxes(v, -1, axis)
    v = nn.Dense(n)(v)
    v = jnp.swapaxes(v, -1, axis)
    return u * (v + 1.)

class SpatialGmlpLayer(nn.Module):
  """Gated MLP applied on a specified axis: -3 for grid and -2 for block."""
  grid_size: Sequence[int]
  block_size: Sequence[int]
  @nn.compact
  def __call__(self, x, axis=-3):
    n, h, w, num_channels = x.shape
    if axis=-3:  # for grid gMLP layer
        gh, gw = self.grid_size
        fh, fw = h // gh, w // gw
    elif axis=-2:  # for block gMLP layer
        fh, fw = self.block_size
        gh, gw = h // fh, w // fw
    x = block_images(x, patch_size=(fh, fw))
    y = nn.LayerNorm()(x)
    y = nn.Dense(num_channels * 2)(y)
    y = nn.gelu(y)
    y = SpatialGatingUnit()(y, axis=axis)
    y = nn.Dense(num_channels)(y)
    x = x + y
    x = unblock_images(x, grid_size=(gh, gw), patch_size=(fh, fw))
    return x

class MultiAxisGmlpBlock(nn.Module):
  block_size: Sequence[int]
  grid_size: Sequence[int]
  @nn.compact
  def __call__(self, x):
    shortcut = x
    n, h, w, num_channels = x.shape
    x = nn.LayerNorm()(x)
    x = nn.Dense(num_channels * 2)(x)
    x = nn.gelu(x)
    # split two heads, then applied grid gMLP and block gMLP respectively.
    u, v = jnp.split(x, 2, axis=-1)
    u = SpatialGmlpLayer(grid_size=self.grid_size)(u, axis=-3)
    v = SpatialGmlpLayer(block_size=self.block_size)(v, axis=-2)
    # Concat and output projection
    x = jnp.concatenate([u, v], axis=-1)
    x = nn.Dense(num_channels)(x)
    x = x + shortcut
    return x
\end{lstlisting}
\end{algorithm*}

\clearpage
\clearpage
{\small
\bibliographystyle{ieee_fullname}
\bibliography{egbib}
}

\end{document}